\documentclass[%
 reprint,
 amsmath,amssymb,
 aps,
]{revtex4-2}

\usepackage{amsmath}
\usepackage{graphicx}
\usepackage{dcolumn}
\usepackage{bm}
\usepackage{xcolor}
\usepackage{diagbox}
\usepackage[bb=boondox]{mathalfa}

\begin{document}
\preprint{APS/123-QED}
\title{Spectral fluctuations and crossovers in multilayer network}

\author{Himanshu Shekhar\textsuperscript{1*}, Ashutosh Dheer\textsuperscript{1}, Santosh Kumar\textsuperscript{1}, N. Sukumar\textsuperscript{2}}

\affiliation{\textsuperscript{1}Department of Physics, Shiv Nadar Institution of Eminence, Gautam Buddha Nagar, Uttar Pradesh 201314, India}
\affiliation{\textsuperscript{2}Amrita School of Artificial Intelligence, Amrita Vishwa Vidyapeetham, Coimbatore, Tamil Nadu, 641112, India.}

\email{\textsuperscript{1}himanshuphy02@gmail.com, \textsuperscript{}ad326@snu.edu.in, \textsuperscript{}Deceased, \textsuperscript{2} n$\_$sukumar@cb.amrita.edu}

\date{\today}

\begin{abstract}

Spectral fluctuation analysis within the random matrix theory (RMT) framework is a powerful probe of complexity in networked systems, yet its extension to multilayer architectures remains unresolved. In general multilayer networks, the adjacency matrix possesses a heterogeneous block structure with unequal variances across layers, causing eigenvalue spacing statistics to deviate from RMT predictions even when each individual layer is perfectly random. We demonstrate that this variance mismatch is the central obstacle to observing spectral universality in multilayer systems and introduce a general block-wise normalization scheme that restores the correct variance structure across all blocks. Using Higher--order spacing ratios, we show that once properly normalized, multilayer networks exhibit universal spectral fluctuations consistent with RMT across a broad class of configurations, including purely intralayer, interlayer, and multiplex structures. Focusing on the bilayer Case, we introduce a crossover model parametrized by the relative interlayer to intralayer coupling strength, which captures the continuous transition from two independent Gaussian orthogonal ensembles (GOEs) to a single GOE. We find that this crossover sharpens with increasing system size, suggesting that in the large-system limit, arbitrarily weak interlayer coupling may be sufficient to induce global spectral correlations. Applying the framework to empirical multilayer networks derived from protein crystal structures, we demonstrate that structural coupling drives analogous spectral transitions, directly linking the emergence of universality to physically meaningful organization. These results establish spectral universality as a robust feature of multilayer networks and provide a quantitative framework for understanding how structure and coupling govern collective behavior in complex interconnected systems.
\end{abstract}


\maketitle

\section{\label{sec:level1}Introduction}

Understanding the structural and dynamical complexity of real-world systems is a central problem in network science. From protein interaction networks and neural circuits to power grids and social systems, interactions among constituent elements give rise to collective behavior~\cite{AL2002, MN2018, SH2001, SV2006, EE2012, WF1994, Larson2021, LH2007, BGL2011, JDB2017, ETYD2018, BJN2002}. To understand structural complexity in networks, synthetic models have been developed within graph theory to provide controlled and analytically tractable frameworks. Canonical network models, including the Erd\H{o}s--R\'{e}nyi (ER) random graph, the Watts--Strogatz small-world network, and the Barab\'{a}si--Albert scale-free network, have served as paradigmatic benchmarks for elucidating how network topology governs dynamical properties such as robustness, diffusion, synchronization, and percolation~\cite{ER1960, Bollobas1998, Gilbert1959, Fienberg2012, WS1998, AB2002}. Spectral analysis provides a systematic framework for probing this complexity: the eigenvalue spectrum of the adjacency matrix encodes structural features such as connectivity patterns and network organization~\cite{RKD2000, Barthelemy2011, Chung1997}. In particular, fluctuations in eigenvalues serve as sensitive indicators of order and randomness in complex systems.

Random matrix theory (RMT) provides the theoretical backbone for such spectral fluctuation analysis. Originally developed to describe energy level statistics in complex quantum systems~\cite{Mehta2004}, RMT predicts that large, disordered systems exhibit universal spectral statistics determined solely by symmetry class, independent of microscopic details. For adjacency matrices of large random graphs, the nearest-neighbor spacing distribution follows the Wigner-Dyson form characteristic of the Gaussian orthogonal ensemble (GOE), a hallmark of chaotic or strongly correlated systems~\cite{BJ2007, JB2007}. The spacing ratio distribution (SRD), introduced as an unfolding-independent alternative, captures the same universal features and is particularly effective for finite-size empirical networks~\cite{OH2007, ABGR2013}. Applications of RMT to complex networks span biological, social, and technological systems~\cite{AA2014, MSN2022, AWJ2022, RL2016}, establishing spectral fluctuation analysis as a robust probe of network complexity.

Most real-world systems, however, cannot be adequately represented as single-layer networks. Multimodal transportation systems, multiplex social networks, interconnected infrastructure grids, and biological assemblies such as protein complexes involve multiple types of connections. The multilayer network formalism~\cite{KABG2014, BBCG2014, DSCK2013, CARM2016, MCMA2016} captures this structure through a block adjacency matrix, where diagonal blocks represent intralayer connections and off-diagonal blocks encode interlayer coupling. This naturally raises fundamental questions about spectral properties: Do multilayer networks retain RMT universality, and how does the interplay between intra- and interlayer connectivity shape spectral fluctuations? Since spectral statistics are closely linked to dynamical processes, addressing these questions is essential for understanding transport, synchronization, and cascading phenomena in interconnected systems.

Existing RMT studies of multilayer networks have largely focused on structurally uniform architectures, where layers share identical node sets and interlayer coupling is represented by identity or scaled identity matrices~\cite{GJFY2017, RJ2022, RJ2023, GM2025}. While these works have established important results on eigenfunction localization~\cite{GJFY2017, GM2025} and spacing ratio statistics~\cite{RJ2022, RJ2023, Martinez2019}, they do not address genuinely heterogeneous multilayer systems, where layers differ in size and connectivity. In such systems, the block variances differ across layers, which leads to systematic deviations from RMT predictions even when individual layers are random. This variance mismatch prevents a consistent characterization of spectral universality in multilayer networks. Developing a general RMT framework that remains valid across arbitrary architectures and resolves this normalization issue is therefore essential.

In this work, we develop a general RMT-based framework for spectral fluctuation analysis in multilayer networks. We consider a broad class of multilayer architectures, including networks with intralayer connections, interlayer connections, and multiplex structures, and employ Higher--order spacing ratios~\cite{SM2020, Bhosale2021, TBS2018, BTS2018} to probe spectral correlations without unfolding. A key contribution is the introduction of block-wise scaling factors that normalize variances across both diagonal and off-diagonal blocks, and remain robust to variations in layer size and connection probabilities. We show that this normalization is essential for recovering RMT universality in heterogeneous multilayer systems.

Our analysis proceeds in two parts. First, we construct ensembles of random multilayer networks with Erd\H{o}s-R\'{e}nyi (ER) connectivity and systematically vary intra- and interlayer connection probabilities across different architectures to explore the dependence of spectral statistics on connectivity patterns. We then introduce a bilayer crossover model parametrized by the relative strength of interlayer to intralayer coupling, which captures the transition in spectral statistics from two independent GOEs to a single GOE. We find that this crossover sharpens with increasing system size, suggesting that in the large-dimension limit, even weak interlayer coupling can drive the system toward single GOE behavior. Second, we apply the framework to empirical multilayer networks derived from protein crystal structures (1EWT, 1EWK, and 1UW6), modeling residues as nodes and spatial proximity as edges. By varying distance thresholds, we observe transitions in spectral statistics consistent with changes in structural coupling. We find that with appropriate block scaling, spectral fluctuations align with RMT predictions even in the presence of structurally distinct layers and varying connectivity, supporting the universality of multilayer network spectra.

The paper is organized as follows.
In Sec~\ref{sec:level2} we define the multilayer network model, its adjacency matrix representation, and the scaling procedure required to recover universal spectral fluctuations. Sec~\ref{sec:level3} presents the spectral statistics of random multilayer networks for different architectures, including the bilayer crossover model. In Sec~\ref{sec:level4} we apply the framework to empirical multilayer networks derived from protein crystal structures, and summarize the results and discuss possible extensions in Sec~\ref{sec:level5}.

\section{\label{sec:level2}Preliminaries}

\subsection{\label{sublevel:level2.1}Multilayer network and general matrix model} 

A multilayer network can be defined as a function $\Phi \equiv \Phi(G,E)$ of two sets: a set of graphs $G$ and a set $E$ of connections. The set $G = \{g_{a}|a=1,2, \ldots, m\}$ is the set of graphs in each layer, where each graph $g_{a} \equiv g_{a}(N_{a}, e_{a})$ is defined by the set of nodes $N_{a}$ and the set of edges $e_{a}$. The set $E = \{e_{ab}; a, b \in \{1,2, \ldots, m\}\}$ contains connections between nodes; elements of $E$ could denote intralayer connections (if $a = b$), or interlayer connections (if $a\neq b$) between nodes $N_{a}$ and $N_{b}$ of layers $g_{a}$ and $g_{b}$~\cite{BBCG2014, RJ2022}.

Networks can be represented by adjacency matrices which for multilayered networks take the form of block-structured matrices. Considering an undirected network structure, an $m$-layer network with each layer having, in general, a different number of nodes ($n_1, n_2, \ldots, n_m$) and can be represented by the following adjacency matrix
\begin{align}\label{eq:1}
\mathcal{A}=\begin{bmatrix}
A_{n_1 \times n_1}^{1} & B_{n_1 \times n_2}^{1,2} & \ldots & B_{n_1 \times n_m}^{1,m} \\[6pt]
(B^{1,2})_{n_2 \times n_1}^{T} & A_{n_2 \times n_2}^{2}& \ldots &  B_{n_2 \times n_m}^{2,m} \\[6pt]
\vdots & \vdots  & \ddots & \vdots \\[6pt]
(B^{1,m})_{n_m \times n_1}^{T} & (B^{2,m})_{n_m \times n_2}^{T}  &\ldots & A_{n_m \times n_m}^{m}\\
\end{bmatrix}.
\end{align}
The $j$th diagonal block, $A^j_{n_j\times n_j}$, is the adjacency matrix of the $j$th layer and is of size $n_j\times n_j$. Its off-diagonal entries carry information about the associated intralayer connections and its diagonal elements are all zero. Hence, within the unweighted network setup, if $a_{xy}^k$ is an element of $A_{n_k \times n_k}^{k}$, then $a_{xy}^k = 0$ implies that nodes $x$ and $y$ in the $k$th layer, are not connected, whereas $a_{xy}^k = 1$ implies that they are connected. $B^{j,k}_{n_j\times n_k}$ are off-diagonal blocks of size $n_j\times n_k$ and represents the interlayer connection between layers $j$ and $k$. If $b_{xy}^{k,l}$ is an element of $B_{n_k \times n_l}^{k,l}$, then $b_{xy}^{k,l} = 0$ means that the node $x$ of the $k$th layer and the node $y$ of the $l$th layer are not connected, whereas $b_{xy}^{k,l} = 1$ implies that they are connected. Since the network is undirected, $\mathcal{A}$ has an overall real-symmetric structure. The overall dimension of $\mathcal{A}$ is $n\times n$ where $n=n_1+n_2+ \ldots +n_m$.

It should be noted that the diagonal blocks are necessarily square, whereas the off-diagonal blocks may be rectangular in general. In Case there are no interlayer connections, then all off-diagonal blocks are zero matrices, and $\mathcal{A}$ is a block-diagonal matrix. The other extreme is that when there is no intralayer connections, then the diagonal blocks would be zero matrices. For multiplex networks, interlayer edges can only connect nodes representing the same entity in different layers. Therefore, $n_1=n_2=\ldots=n_m$ and the off-diagonal blocks would be identity matrices in this Case.

Finally, instead of binary values of $0$ and $1$ for the elements, we may consider any real number from the interval $[0,1]$, in which Case we have an adjacency matrix corresponding to a weighted network.

\subsection{\label{sublevel:level2.2}Spectral fluctuation properties in RMT}

Spectral fluctuations of the eigenvalues of an operator associated with a system encode important information about its underlying dynamics. The agreement of these fluctuations with the predictions of RMT serves as an indicator of chaotic or complex behavior of the system. Among the various measures for quantifying short-range spectral fluctuations, the distribution of spacing ratios stands out as a widely used measure. Its appeal lies in the fact that it does not require prior knowledge of the local density of states, thereby circumventing the need for spectral unfolding~\cite{OH2007,ABGR2013}. 

Let us consider the ordered set of eigenvalues $\{\lambda_{1}, \lambda_{2},  ...., \lambda_{n}\}$ obtained from diagonalizing a given matrix. The consecutive spacing ratios are then defined as
\begin{align}
\label{ratio1}
r_{j} = \frac{\lambda_{j+2}-\lambda_{j+1}}{\lambda_{j+1}-\lambda_{j}},~~~  j = 1, 2, ...., n-2.
\end{align}
The distribution of these spacing ratios, called the spacing ratio distribution (SRD), is known~\cite{OH2007,ABGR2013} for the classical Gaussian random matrix ensembles and is given by
\begin{eqnarray}\label{WDsurmise}
P(\alpha,r)=C_{\alpha} \frac{(r+r^2)^\alpha}{(1+r+r^2)^{1+3 \alpha/2}},
\end{eqnarray}
where $C_{\alpha}$ is the normalization factor such that $\int_0^\infty P(\alpha,r)dr=1$. The parameter $\alpha$ in the above expression is related to the symmetry properties of the random matrix under consideration and coincides with the Dyson index ($\beta$) for the classical random matrix ensembles~\cite{Mehta2004}. The Gaussian orthogonal ensemble (GOE) consists of real-symmetric random matrices whose independent entries are Gaussian distributed with zero mean, where the off-diagonal elements have variance $\sigma^2$ and the diagonal elements have variance $2\sigma^2$. It describes systems with time-reversal symmetry. The Gaussian unitary ensemble (GUE) and Gaussian symplectic ensemble (GSE) correspond to complex-Hermitian and quaternion self-dual matrices, respectively, associated with systems without time-reversal symmetry and with time-reversal symmetry including spin. In these Cases, $\alpha$ assumes the values $1$, $2$, and $4$ for the GOE, GUE, and GSE, respectively, reflecting the number of independent degrees of freedom per matrix element in real-symmetric, complex-Hermitian, and quaternion self-dual ensembles. Remarkably, by appropriately fixing the parameter $\alpha$, the above formula has also been found to work for Higher--order spacing ratios, as well as in the Case of matrices having a block-diagonal structure.

Higher--order spacing ratios, beyond the nearest-neighbor ratio defined in Eq.~\eqref{ratio1}, have been investigated and applied in various contexts such as quantum chaos, many-body non-integrability, and Floquet dynamics~\cite {SM2020, Bhosale2021, ONF2022, Ankit2024}. A notable example is that of $k$-th order (non-overlapping) spacing ratio, defined as~\cite{BTS2018,TBS2018, SM2020, Bhosale2021}
\begin{equation}\label{ratio2}
r^{(k)}_i=\frac{\lambda_{i+2k}-\lambda_{i+k}}{\lambda_{i+k}-\lambda_{i}},\hspace{1cm} i,k=1,2,3 \dots.
\end{equation}
For $k=1$, one recovers Eq.~\eqref{ratio1}. In Ref.~\cite{Bhosale2021}, the $k$-th order spacing ratio has been used to study the superposition of random matrix spectra. One can also consider block-diagonal random matrices, where each block corresponds to a standard classical ensemble, e.g., a direct sum of two GOE matrices. In such Cases, an appropriate value of $\alpha$ can be identified~\cite{Bhosale2021} such that the $k$-th order spacing ratio distribution of the combined spectrum follows Eq.~\eqref{WDsurmise}. In the present work, we focus on real-symmetric matrices and list the corresponding $\alpha$ values for a given order $k$ and number of diagonal blocks $m$ in Table~\ref{table:1}.

\begin{table}
\centering
\begin{tabular}{|l||*{5}{c|}}\hline
\backslashbox{$k$}{$m$}
&\makebox[3em]{1}&\makebox[3em]{2}&\makebox[3em]{3}&\makebox[3em]{4}\\ \hline\hline
1 &1& $\sim 0$ & $\sim 0$ &$\sim 0$\\ \hline
2 &4&2& $\sim 1.25$&1 \\ \hline
3 &8& $ 4$ &3&$\sim 2.5$ \\ \hline
4 &13&7&5&4\\ \hline
\end{tabular}
\caption{Tabulation of Higher--order indices $\alpha$ for various k and superposition of m GOEs.}
\label{table:1}
\end{table}

For the block-diagonal construction to exhibit spectral fluctuations consistent with Eq.~\eqref{WDsurmise} for a given value of $\alpha$ as listed in Table~\ref{table:1}, the diagonal blocks must have the same variance scale. This can be expressed by requiring that the quantity $\langle \mathrm{tr}(A^{(j)})^2 \rangle/(n_j+1)$ be identical across all blocks $A^{(j)}_{n_j \times n_j}$, where $\langle \cdot \rangle$ denotes ensemble averaging~\cite{Mehta2004}. If this condition is not satisfied, the blocks must be appropriately rescaled. For example, in a two-block Case with distributions $P(A^{(j)}) \propto \exp\left[-\frac{1}{2\sigma_j^2} \mathrm{tr}(A^{(j)})^2\right]$, $j=1,2$, one has $\langle \mathrm{tr}(A^{(j)})^2 \rangle/(n_j+1) = n_j \sigma_j^2/2$. The rescaled matrices are then given by
\begin{align*}
\widetilde{A}^{(j)} = \frac{a\, A^{(j)}}{(n_j \sigma_j^2/2)^{1/2}},
\end{align*}
where $a$ is an arbitrary nonzero constant~\cite{Farkas2001, Zhao2012}. A similar normalization is required when constructing adjacency matrices from Erd\H{o}s-R\'enyi random graphs, as discussed in the next section.

\section{\label{sec:level3}Spectral Fluctuations of random multilayer network}

Random networks arise in a wide range of complex systems, including social, biological, and technological networks~\cite{Hammoud2020,AM2019,JM2018,Goldblum2019,Jalan2014,AJalan2014}. Despite differences in architecture and connectivity patterns, such systems often exhibit universal spectral properties well described by RMT~\cite{JB2007,BJ2007}. In this section, we investigate spectral fluctuations in multilayer random networks, focusing on how variations in connection patterns across and within layers influence these fluctuations, and whether the universal behavior predicted by RMT survives in the presence of multiple interacting layers.

For our analyses, we construct blocks of the overall adjacency matrix using the Erd\H{o}s-R\'enyi (ER) approach. The $j$th diagonal block $A_{n_j\times n_j}^j$ is constructed using the usual Erd\H{o}s-R\'enyi $G(n_j,p_j)$ graph model in which the $n_j$ nodes (vertices) are connected randomly in a manner such that each edge is included in the graph with probability $p_j$, independently from every other edge. Consequently, a graph in $G(n_j,p_j)$ has on an average $n_j(n_j-1)p_j/2$ edges. To keep the variances comparable, we multiply the $j$th block by a factor of $1/\sqrt{4n_{j}p_{j}(1-p_{j})}$. This ensures that $\langle \mathrm{tr}(A_j^2) \rangle$ scales consistently across blocks. In this way, the diagonal blocks retain similar spectral fluctuations, ensuring agreement with the results of RMT. Without this scaling, the resulting spectral fluctuations deviate from RMT predictions because of the statistical mismatch across layers.

An off-diagonal block $B_{n_j\times n_k}^{j,k}$ is also constructed in a similar manner by connecting the $n_j$ nodes of the $j$th layer with the $n_k$ nodes of the $k$th with probability $p_{jk}$, $j\ne k$. Thus, on average, there will be $n_j n_k p_{jk}$ edges between the two layers.
To ensure comparable variance across all blocks, we scale the off-diagonal blocks by the factor $1/\sqrt{4\sqrt{n_{j}n_{k}}p_{jk}(1-p_{jk})}$.

\begin{figure*}[!ht]
 \centering
 \includegraphics[width=0.7\textwidth]{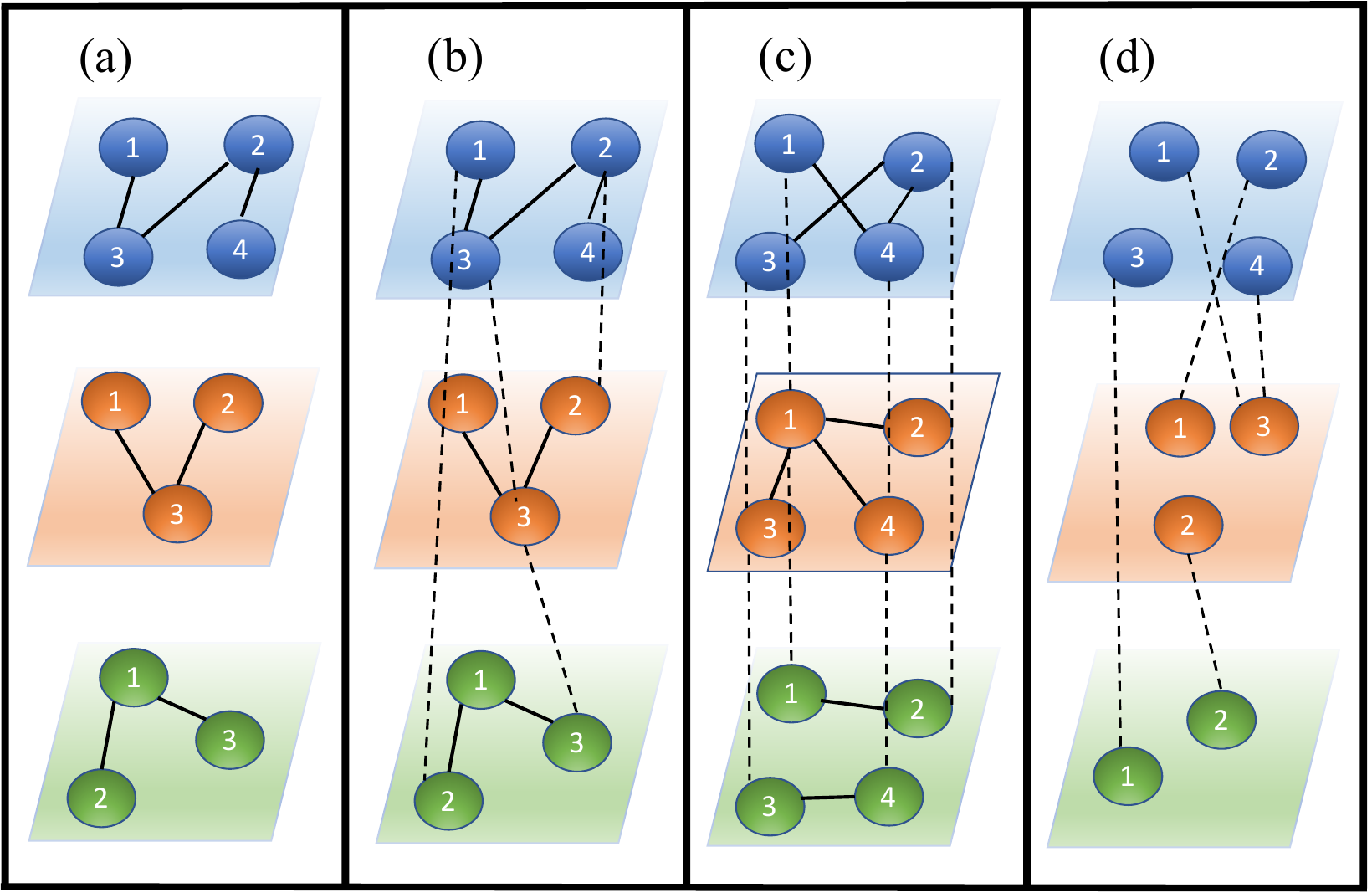}
 \caption{Schematic diagram representing examples of multilayer networks. The black dashed line indicates an interlayer connection, while the solid black line depicts an intralayer connection. (a) A multilayer network with only intralayer connections, no interlayer connections. (b) A general multilayer network where interlayer edges connect different nodes across the layers. (c) A multiplex network where the interlayer edges connect the same nodes across the layers. (d) Multiparite network with only interlayer connections, no intralayer connections.}
 \label{fig:multilayer}
\end{figure*}

We consider the following four different types of possibilities, as also depicted in Fig.~\ref{fig:multilayer}.

\vspace{0.2cm}
\noindent
{\it Case a.} In this Case, there are no connections between nodes of different layers ($p_{jk}=0$ for all $j\ne k$) and therefore all off-diagonal blocks in the adjacency matrix $\mathcal{A}$ are zero matrices. The diagonal blocks associated with the intralayer connections have random entries $0$ or $1$ as decided by the Erd\H{o}s-R\'enyi construction. The $\mathcal{A}$ matrix in this Case, therefore, assumes a block-diagonal form. 

\vspace{0.2cm}
\noindent
{\it Case b.} In this Case, we consider both intralayer and interlayer connections to be random, so that the adjacency matrix of the multilayer network is fully random. 

\vspace{0.2cm}
\noindent
{\it Case c.} This is the Case corresponding to a multiplex network, for which interlayer edges connect the same nodes across all the layers, so all the off-diagonal blocks of $\mathcal{A}$ are identity matrices. The randomness arises solely from the diagonal blocks.

\vspace{0.2cm}
\noindent
{\it Case d.} Here, we consider no intralayer connections ($p_j=0$ for all $j$). Consequently, all diagonal blocks in $\mathcal{A}$ are zero matrices, and the off-diagonal blocks for interlayer connections contain random entries taking values 0 or 1.

\begin{figure*}[!ht]
    \centering
    \includegraphics[width=0.8\textwidth]{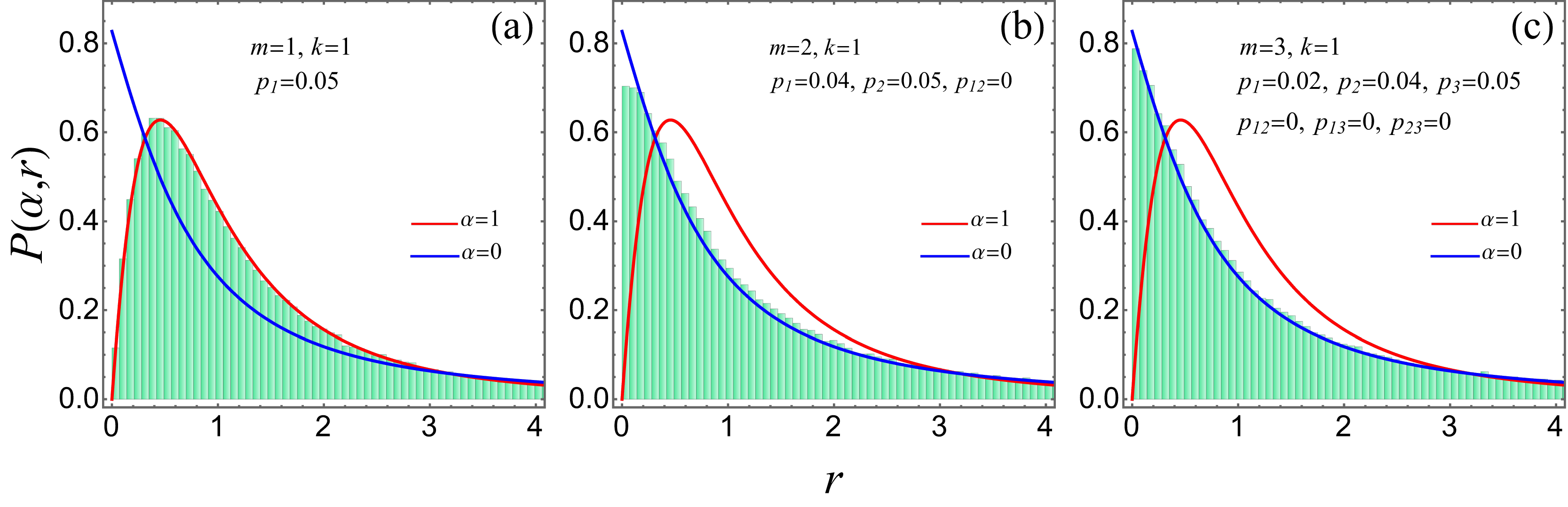}
    \caption{The first-order ($k=1$) SRD plots (histogram) is obtained through simulations based on an ensemble of 250 adjacency matrices $\mathcal{A}$ for three network configurations. (a) A single-layer network with dimensions $n_1 = 800 \times 800$ and an edge connection probability of $p_1 = 0.05$; (b) A bilayer network comprising block-diagonal matrices of dimensions $n_1 = 350 \times 350$ and $n_2 = 450 \times 450$, with edge connection probabilities $p_1 = 0.04$, $p_2 = 0.05$, and no interlayer connections ($p_{12} = 0$). (c) A trilayer network with block-diagonal matrices of dimensions $200 \times 200$, $300 \times 300$, and $500 \times 500$, with edge connection probabilities $p_1 = 0.02$, $p_2 = 0.04$, $p_3 = 0.05$, and no interlayer connections ($p_{12} = 0$, $p_{13} = 0$, $p_{23} = 0$).}
    \label{fig:first_Order_flucuation}
\end{figure*}

\vspace{0.2cm}
\noindent
In this study, for simplicity, we focus on bilayer ($m=2$) and trilayer ($m=3$) networks to examine their spectral fluctuations, particularly for Higher--order fluctuations where $k \geq 2$. For $k=1$, only single-layer networks ($m=1$) exhibit a correlated spacing ratio distribution (SRD), while bilayer and trilayer networks typically display an uncorrelated spacing ratio distribution, as illustrated in Fig.~\ref{fig:first_Order_flucuation}, which also conforms to Table~\ref{table:1}. Thus, to ensure consistency, we only use Higher--order spectral fluctuation measurements across all network configurations.

\begin{figure*}[!ht]
    \centering
    \includegraphics[width=0.6\textwidth]{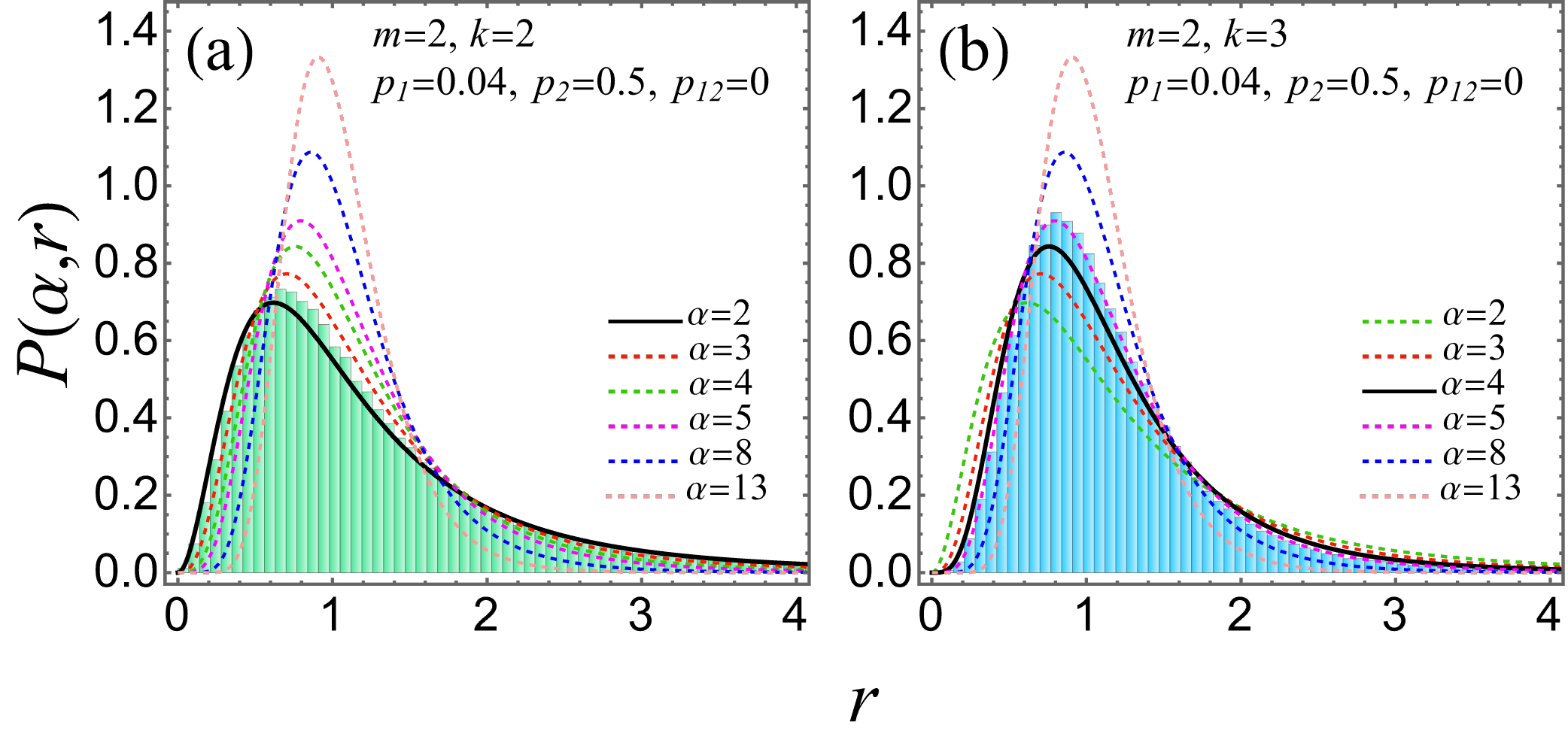}
    \caption{Higher--order SRD plots (histogram) for bilayer network ($m=2$) with no interlayer connections (Case a) without scaling of the two blocks. The connection probabilities being very different in the two layers lead to outcomes different from RMT predictions. The diagonal blocks of the adjacency matrix $\mathcal{A}$ are of dimension $350 \times 350$ and $450 \times 450$, respectively. The histograms have been obtained using simulations based on an ensemble comprising 250 adjacency matrices. The order of the spacing distribution and edge probabilities, $(k,p_1,p_2)$ for the histograms obtained from numerical simulations are (a) $(2,0.04,0.5)$, (b) (3,0.04,0.5).}
    \label{fig:unscaled_blockform_350450}
\end{figure*}

Before analyzing the different multilayer configurations, it is instructive to examine the consequences of not scaling the diagonal and the off-diagonal blocks discussed earlier. We consider a bilayer network ($m=2$) under {\it Case a}, where the two layers have significantly different connection probabilities. Without scaling, the eigenvalue distributions yield spacing ratio statistics that deviate from RMT expectations and do not correspond to any well-defined $\alpha$ value. This is evident in Fig.~\ref{fig:unscaled_blockform_350450}, where the second-order ($k=2$) and third-order ($k=3$) spacing ratio distributions fail to align with Eq.~\eqref{WDsurmise} for any $\alpha$. Upon applying the appropriate scaling, as we will see in the next subsection, the distributions exhibit excellent agreement with RMT predictions for $\alpha = 2$ and $\alpha = 4$, respectively. Hereafter, we always implement the appropriate scaling for each block (diagonal and off-diagonal) to examine the fluctuation properties.

\subsection{\label{sublevel:level3.1} Multilayer network without interlayer connections}
\begin{figure*}[!ht]
    \centering
    \includegraphics[width=0.95\textwidth]{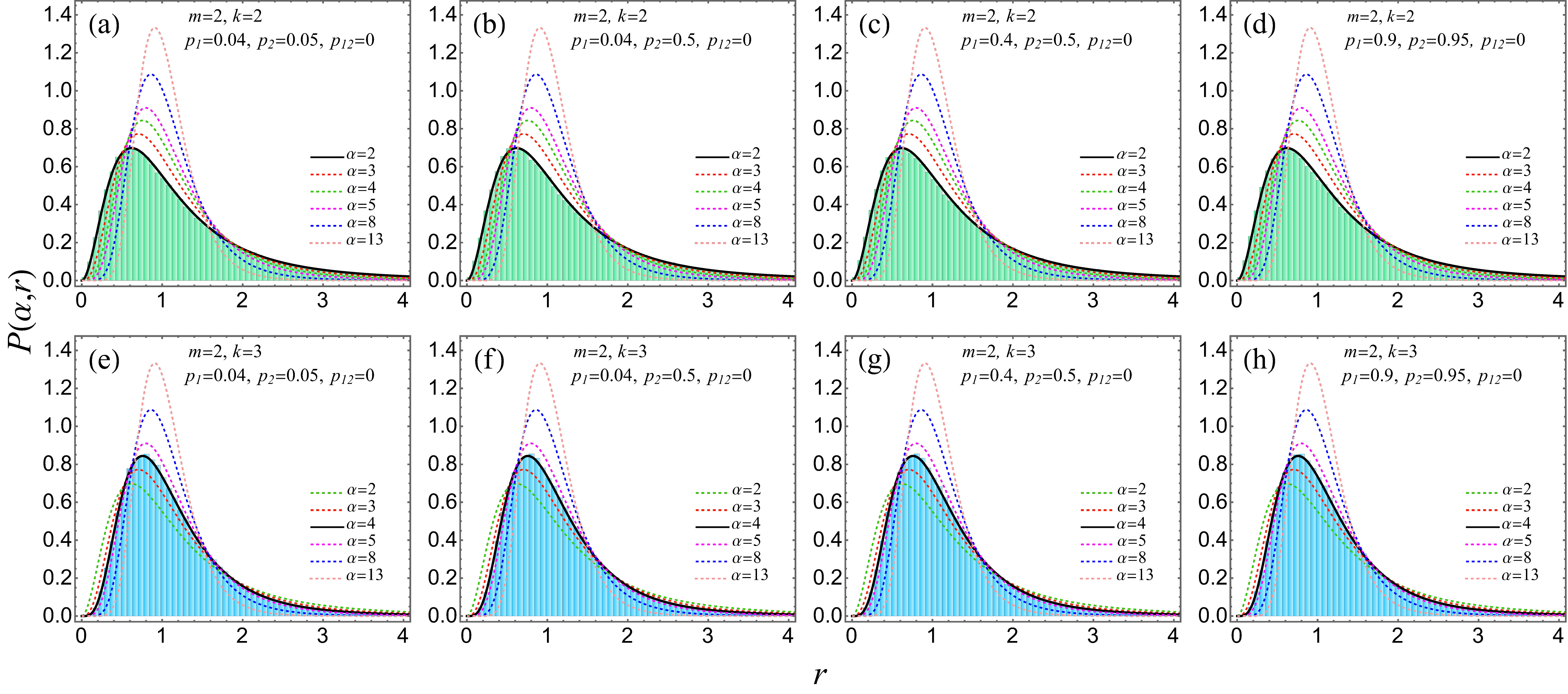}
    \caption{Higher--order SRD plots (histogram) for bilayer network ($m=2$) with no interlayer connections (Case a). The diagonal blocks of the adjacency matrix $\mathcal{A}$ are of dimension $350 \times 350$ and $450 \times 450$, respectively. The histograms have been obtained using simulations based on an ensemble comprising 250 adjacency matrices. The order of the spacing distribution and edge probabilities, $(k,p_1,p_2)$ for the histograms obtained from numerical simulations are (a) (2,0.04,0.05), (b) (2,0.04,0.5), (c) (2,0.4,0.5), (d) (2,0.9,0.95), (e) (3,0.04,0.05), (f) (3,0.04,0.5), (g) (3,0.4,0.5), and (h) (3,0.9,0.95).}
    \label{fig:blockform_350450}
\end{figure*}

\begin{figure*}[!ht]
    \centering
    \includegraphics[width=0.95\textwidth]{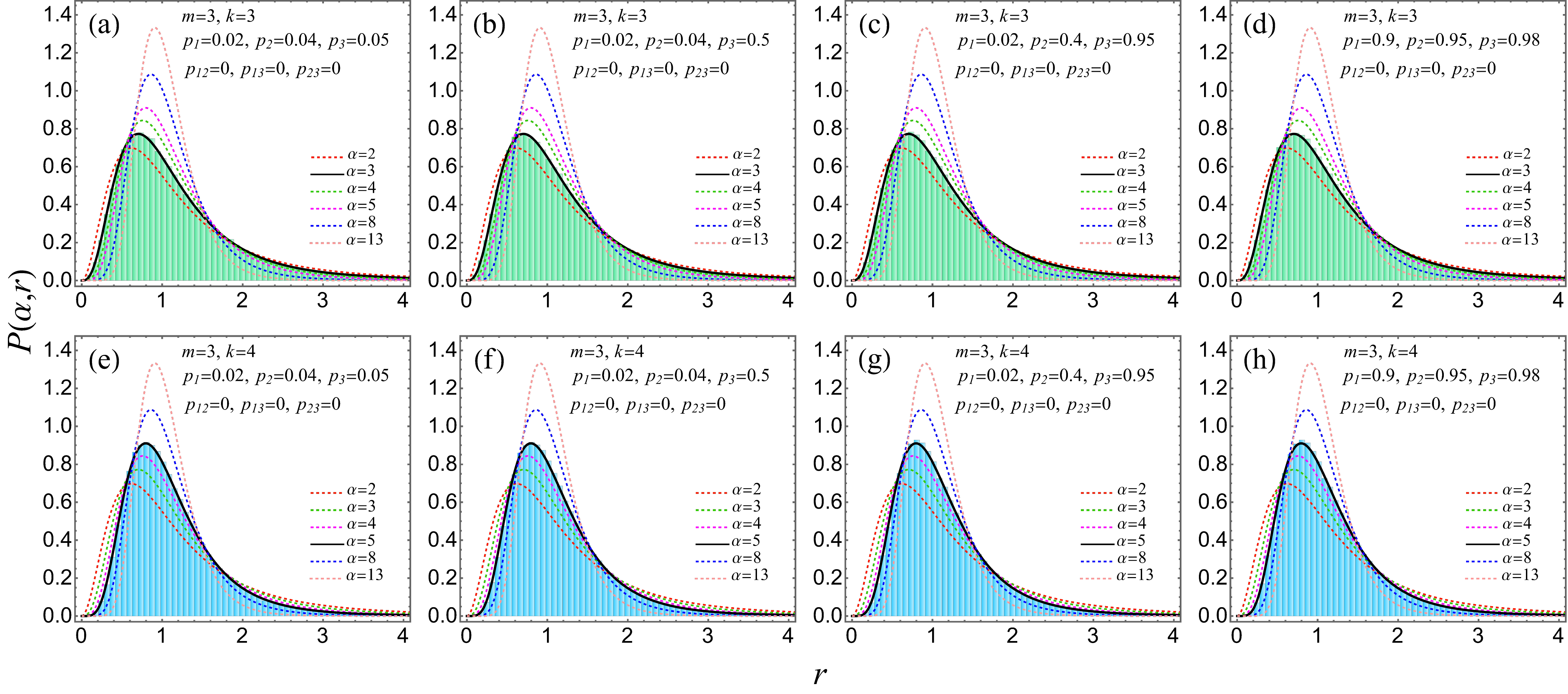}
    \caption{Higher--order  SRD plots (histogram) for trilayer network ($m=3$) with no interlayer connections (Case a). The diagonal blocks of the adjacency matrix $\mathcal{A}$ are of dimension $200 \times 200$, $300 \times 300$ and $500 \times 500$, respectively. The histograms have been obtained using simulations based on an ensemble comprising 200 adjacency matrices. The order of the spacing distribution and edge probabilities $(k,p_1,p_2,p_3)$ for the histograms obtained from numerical simulations are (a) (3,0.02,0.04,0.05), (b) (3,0.02,0.04,0.5), (c) (3,0.02,0.4,0.95), (d) (3,0.9,0.95,0.98), (e) (4,0.02,0.04,0.05), (f) (4,0.02,0.04,0.5), (g) (4,0.02,0.4,0.95), and (h) (4,0.9,0.95,0.98).}
    \label{fig:blockform_200300500}
\end{figure*}

This corresponds to {\it Case a}, where all the off-diagonal blocks of the adjacency matrix $\mathcal{A}$ are zero matrices.
We examine spectral fluctuations for various choices of the edge connection probability $p_j$ in the diagonal blocks to test the universality of the spectral statistics.
In RMT terminology, this Case would correspond to the direct sum of $m$ GOE matrices for an $m$ layer network~\cite{Bhosale2021}.
By appropriately scaling the $j$th block, we ensure that the SRD aligns with the RMT predictions, as illustrated in Figs.~\ref{fig:blockform_350450} and \ref{fig:blockform_200300500}.

In Fig.~\ref{fig:blockform_350450}, we observe that for a bilayer block structure, the distributions correspond to $\alpha = 2$ and $\alpha = 4$ for $k = 2$ and $k = 3$, respectively, across different combinations of edge connection probability $p_j$ in the layers. Similarly, Fig.~\ref{fig:blockform_200300500} illustrates same  for a trilayer block structure, the distributions correspond to $\alpha = 3$ and $\alpha = 5$ for $k = 3$ and $k = 4$, respectively, under various combinations of edge connection probability $p_j$ in the layers.

\subsection{\label{sublevel:level3.2} Multilayer network  with both interlayer and intralayer connections}

In this setup, all blocks of the adjacency matrix $\mathcal{A}$ are connected with edges defined by specific connection probabilities; this aligns with {\it Case b}. Spectral fluctuations are analyzed by exploring various edge connection probabilities $p_j$ for diagonal blocks and $p_{jk}$ for off-diagonal blocks. In the context of RMT, this scenario exhibits spectral fluctuations consistent with those of a single GOE for an $m$-layer network, at the level of spacing ratio statistics \cite{RJ2023}.

\begin{figure*}[!ht]
    \centering
    \includegraphics[width=0.95\textwidth]{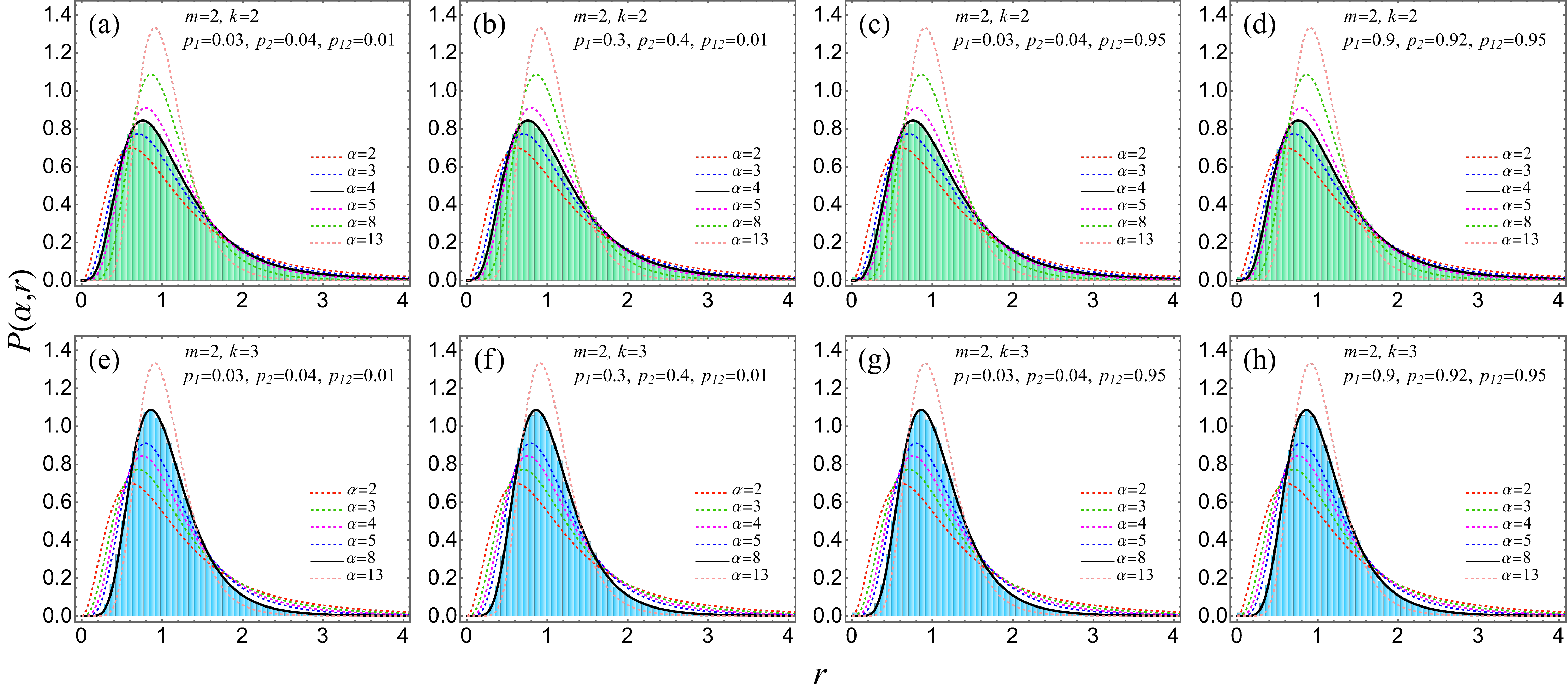}
    \caption{Higher--order SRD plots (histogram) for bilayer network ($m=2$) with  intralayer and interlayer  connections (Case b). The diagonal blocks of the adjacency matrix $\mathcal{A}$ are of dimension $350 \times 350$ and $450 \times 450$, respectively. The histograms have been obtained using simulations based on an ensemble comprising 250 adjacency matrices. The order of the spacing distribution and edge probabilities $(k,p_1,p_2,p_{12})$ for the histograms obtained from numerical simulations are (a) (2,0.03,0.04,0.01), (b) (2,0.3,0.4,0.01), (c) (2,0.03,0.04,0.95), (d) (2,0.9,0.92,0.95), (e) (3,0.03,0.04,0.01), (f) (3,0.3,0.4,0.01), (g) (3,0.03,0.04,0.95), and (h) (3,0.9,0.92,0.95)}
    \label{fig:Bothconnections350_450}
\end{figure*}

\begin{figure*}[!ht]
    \centering
    \includegraphics[width=0.95\textwidth]{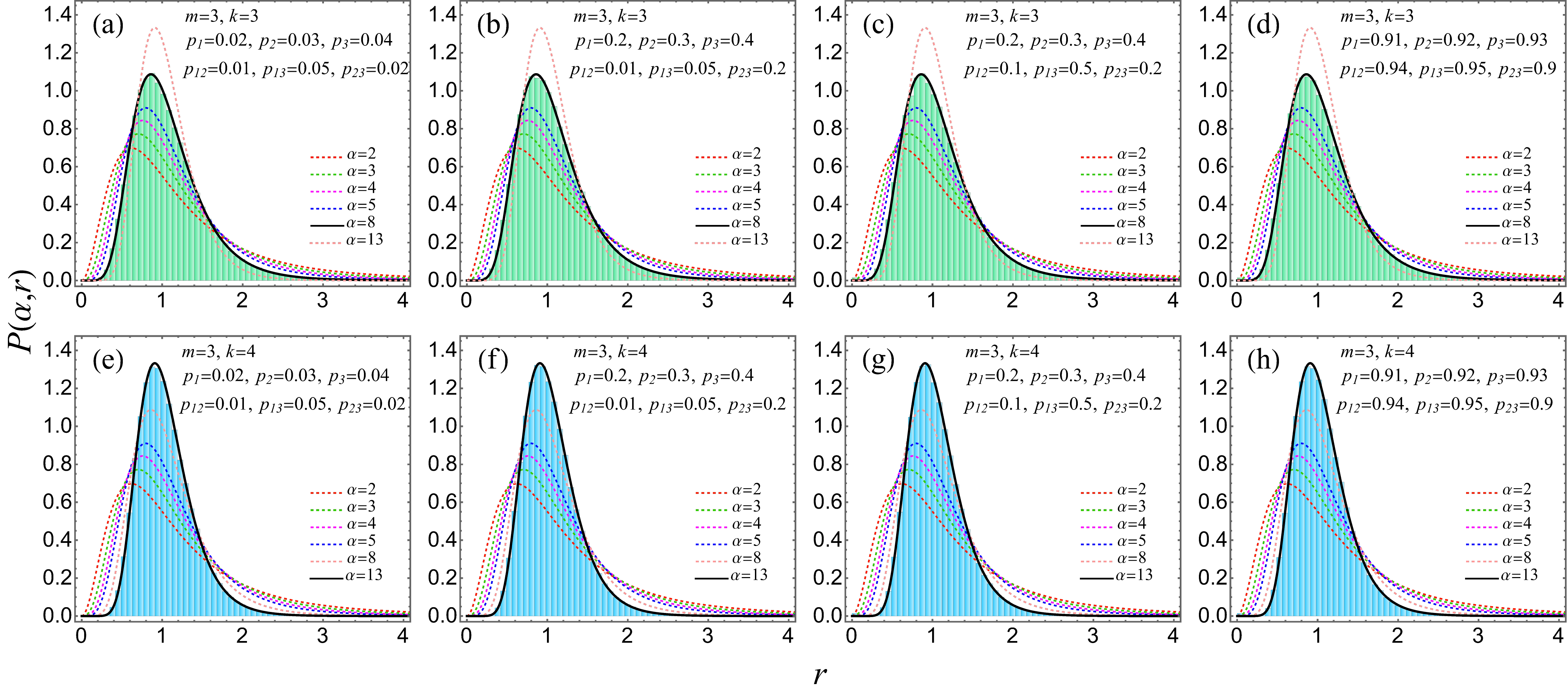}
    \caption{Higher--order  SRD plots (histogram) for trilayer network ($m=3$) with intralayer and interlayer  connections (Case b). The diagonal blocks of the adjacency matrix $\mathcal{A}$ are of dimension $200 \times 200$, $300 \times 300$ and $500 \times 500$, respectively. The histograms have been obtained using simulations based on an ensemble comprising 200 adjacency matrices. The order of the spacing distribution and edge probabilities $(k,p_1,p_2,p_3,p_{12},p_{13},p_{23})$ for the histograms obtained from numerical simulations are (a) (3,0.02,0.03,0.04,0.01,0.05,0.02), (b) (3,0.2,0.3,0.4,0.01,0.05,0.2), (c) (3,0.2,0.3,0.4,0.9,0.5,0.2), (d) (3,0.91,0.92,0.93,0.95,0.95,0.9), (e) (4,0.02,0.03,0.04,0.01,0.05,0.02), (f) (4,0.2,0.3,0.4,0.01,0.05,0.2), (g) (4,0.2,0.3,0.4,0.9,0.5,0.2), and (h) (4,0.91,0.92,0.93,0.95,0.95,0.9)}
    \label{fig:Bothlconnections200_300_500}
\end{figure*}

We analyze SRD for a bilayer network with $k = 2$ and $k = 3$, finding that they correspond to $\alpha = 4$ and $\alpha = 8$, respectively, as shown in Fig.~\ref{fig:Bothconnections350_450}. As indicated in Table~\ref{table:1}, these $\alpha$ and $k$ values are representative of a single-layer network ($m=1$). Similarly, for a trilayer network with $k = 3$ and $k = 4$, the results correspond to $\alpha = 8$ and $\alpha = 13$, respectively, as illustrated in Fig.~\ref{fig:Bothlconnections200_300_500}, aligning once more with the behavior expected for a single-layer network, as per Table~\ref{table:1}.

Next, we investigate the adjacency matrix $\mathcal{A}$ for a multiplex network, {\it Case c}, where the off-diagonal blocks are identity matrices, and randomness comes from different edge connection probabilitoes ($p_j$) of intralayer connections. In this Case too, we find that the SRD for the bilayer network with $k = 2$ and $k = 3$ corresponds to $\alpha = 4$ and $\alpha = 8$, respectively. Similarly, for the trilayer network with $k = 3$ and $k = 4$, the SRD values correspond to $\alpha = 8$ and $\alpha = 13$, respectively. These findings also indicate that the spectral fluctuations align with that of a single-layer network, as indicated in Table~\ref{table:1}.

\subsection{\label{sublevel:level3.3} Multilayer network  with no intralayer connections}

This is {\it Case d}, where all diagonal blocks of the adjacency matrix $\mathcal{A}$ are zero matrices. We examine spectral fluctuations by considering various choices of the edge connection probability $p_{jk}$ for the off-diagonal blocks. In the context of RMT, this scenario also exhibits spectral fluctuations consistent with those of a single GOE for an $m$-layer network, at the level of spacing ratio statistics \cite{Martinez2019}.

\begin{figure*}[!ht]
    \centering
    \includegraphics[width=0.95\textwidth]{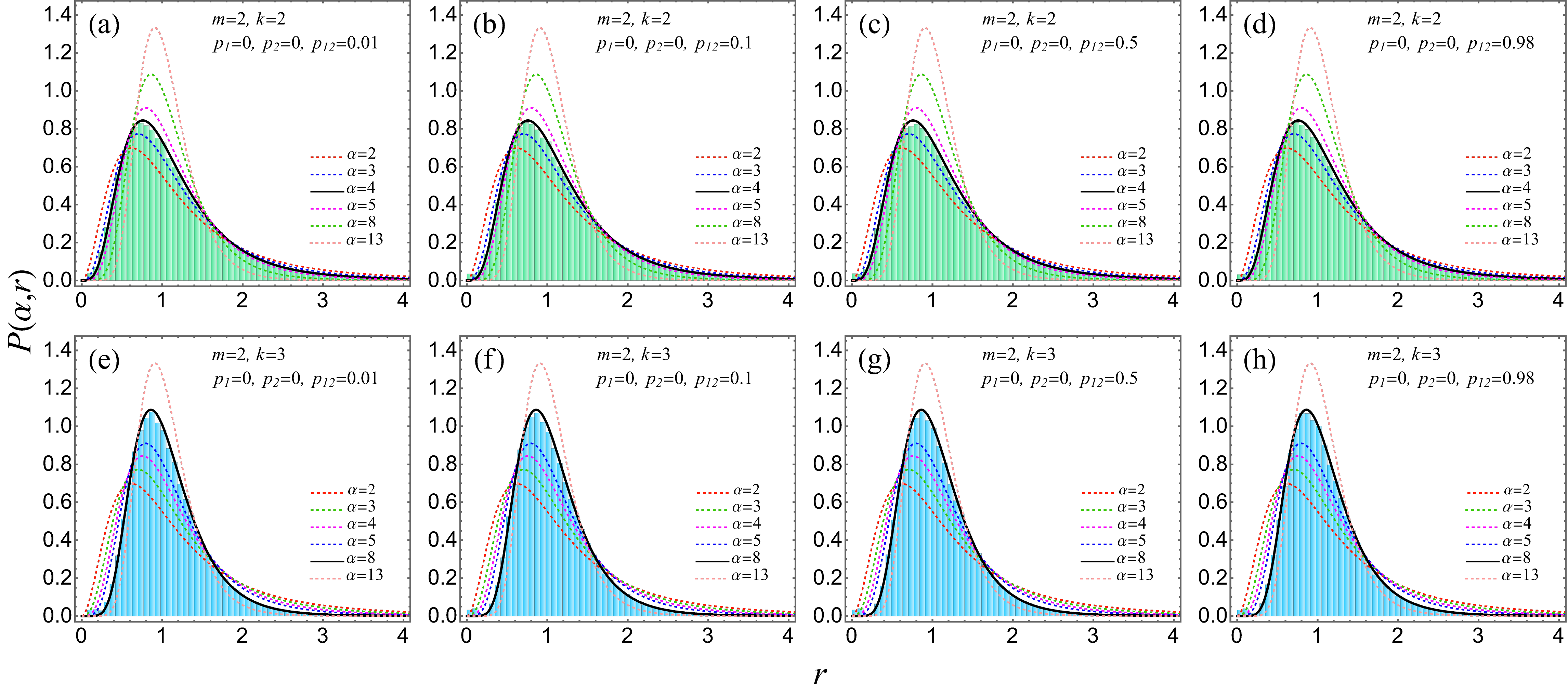}
    \caption{Higher order SRD plots (histogram) for bilayer network ($m=2$) with no intralayer connections (Case d). The diagonal blocks of the adjacency matrix $\mathcal{A}$ are of dimension $350 \times 350$ and $450 \times 450$, respectively. The histograms have been obtained using simulations based on an ensemble comprising 250 adjacency matrices. The order of the spacing distribution and edge probabilities, $(k,p_{12})$ for the histograms obtained from numerical simulations are (a) (2,0.01), (b) (2,0.1), (c) (2,0.5), (d) (2,0.98), (e) (3,0.01), (f) (3,0.1), (g) (3,0.5), and (h) (3,0.98)}
    \label{fig:Interlayer350_4500}
\end{figure*}

\begin{figure*}[!ht]
    \centering
    \includegraphics[width=0.95\textwidth]{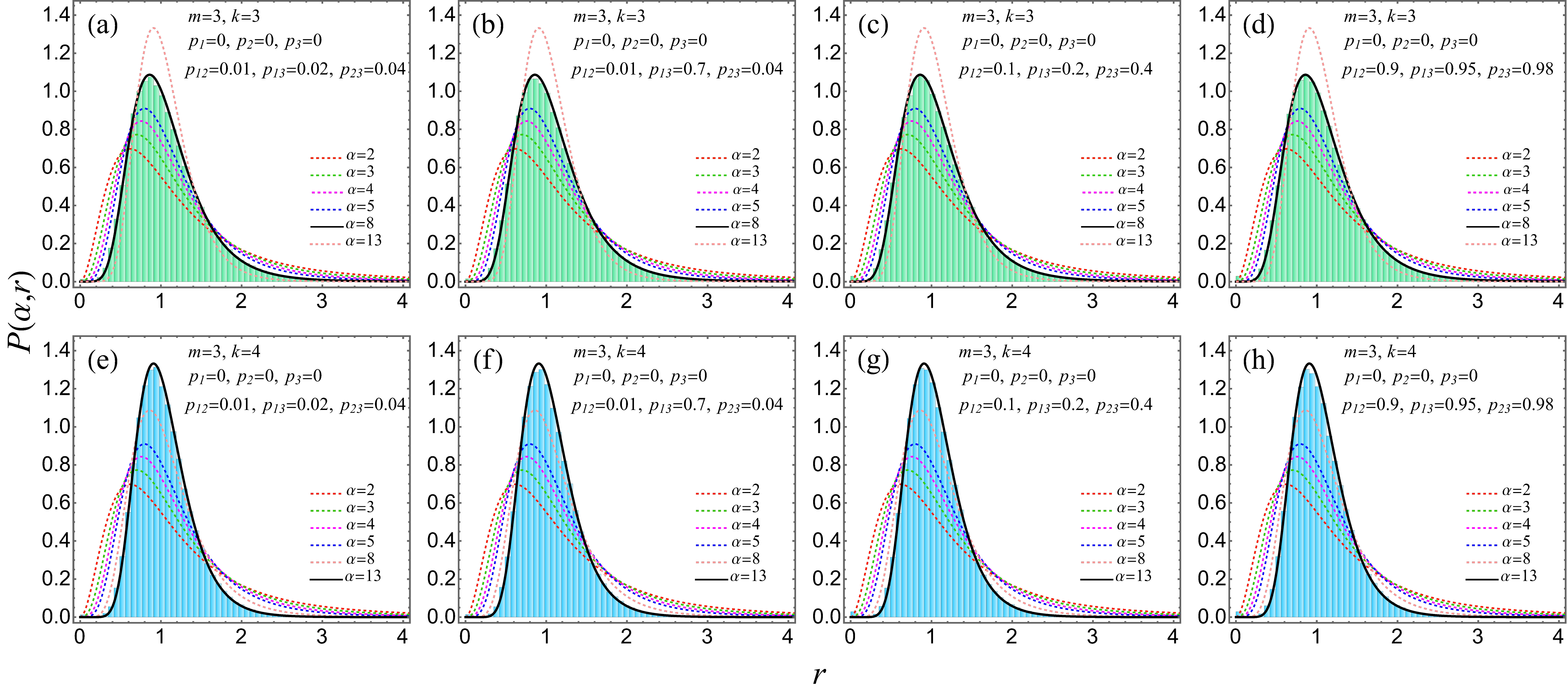}
    \caption{Higher order  SRD plots (histogram) for trilayer network ($m=3$) with no intralayer connections (Case d). The diagonal blocks of the adjacency matrix $\mathcal{A}$ are of dimension $200 \times 200$, $300 \times 300$ and $500 \times 500$, respectively. The histograms have been obtained using simulations based on an ensemble comprising 200 adjacency matrices. The order of the spacing distribution and edge probabilities, $(k,p_{12},p_{13},p_{23})$ for the histograms obtained from numerical simulations are (a) (3,0.01,0.02,0.04), (b) (3,0.01,0.7,0.04), (c) (3,0.1,0.2,0.4), (d) (3,0.9,0.95,0.98), (e) (4,0.01,0.02,0.04), (f) (4,0.01,0.7,0.04), (g) (4,0.1,0.2,0.4), and (h) (4,0.9,0.95,0.98)}
    \label{fig:Interayer200_300_500}
\end{figure*}

We examined the SRD for the bilayer network with $k = 2$ and $k = 3$, finding that they correspond to $\alpha = 4$ and $\alpha = 8$, respectively, as shown in Fig.~\ref{fig:Interlayer350_4500}, which aligns with the single-layer network Case, as indicated in Table~\ref{table:1}. Similarly, for the trilayer network with $k = 3$ and $k = 4$, the SRD values match with $\alpha = 8$ and $\alpha = 13$, respectively, as illustrated in Fig.~\ref{fig:Interayer200_300_500}, and they also correspond to the single-layer network Case, as indicated by Table~\ref{table:1}.

\vspace{10pt}
\subsection{\label{sublevel:level3.4} A crossover model for the bilayer network} 

In the previous three subsections (Secs.  \ref{sublevel:level3.1}-\ref{sublevel:level3.3}), we demonstrated that the Higher--order SRD of the (scaled) eigenvalues of the adjacency matrix for different types of multilayer networks is in remarkable agreement with the predictions from RMT. To further investigate spectral fluctuations through SRD in a bilayer network, we introduce a crossover model defined as follows
\begin{widetext}  
\begin{eqnarray}\label{crossoverequation}
  \mathcal{A}=(1-\gamma)
  \begin{bmatrix}
    A_{n_1 \times n_1}^{1} & 0 \\
  0 & A_{n_2 \times n_2}^{2}  \\
  \end{bmatrix} +
 \gamma \begin{bmatrix}
  0 & B_{n_1 \times n_2}^{1,2} \\
  (B^{1,2})_{n_2 \times n_1}^{T}  & 0  \\
  \end{bmatrix}.
\end{eqnarray}
\end{widetext}  
Here, $\gamma$ is a crossover parameter, $A_{n_1 \times n_1}^{1}$ and $A_{n_2 \times n_2}^{2}$ are the block matrices corresponding to intralayer connections of layers 1 and 2, respectively. The matrix $B_{n_1 \times n_2}^{1,2}$ represents interlayer connections between the two layers, and $(B^{1,2})_{n_2 \times n_1}^{T}$ is the transpose of $B_{n_1 \times n_2}^{1,2}$. The blocks $A_{n_1 \times n_1}^{1}$, $A_{n_2 \times n_2}^{2}$, and $B_{n_1 \times n_2}^{1,2}$ are constructed from the ER network, with edge connection probabilities $p_{1}$, $p_{2}$, and $p_{12}$, respectively. The parameter $\gamma$ can vary from 0 to 1, with the extremes corresponding to only intralayer connections ($\gamma=0$) and only interlayer connections ($\gamma=1$). When $\gamma=0$, the model reduces to a block diagonal structure, and the spectral fluctuations correspond to those of two independent GOE ensembles for the $m=2$ layer network, as seen in Sec. ~\ref{sublevel:level3.1}. When $\gamma=1$, the spectral fluctuations correspond to those of a single GOE, as observed in Sec. ~\ref{sublevel:level3.3}. This crossover model thus provides a simple framework to interpolate between two limiting regimes of multilayer networks: independent layers and fully coupled systems. For small values of $\gamma$, the layers are weakly connected and behave approximately as independent subsystems, whereas for larger $\gamma$, interlayer coupling induces strong mixing between layers, leading to collective behavior. Thus, the parameter $\gamma$ controls the degree of structural coupling and essentially serves as a strength factor for the off-diagonal blocks (or interlayer connectivity); as $\gamma$ increases, the relative contribution of the off-diagonal blocks increases while that of the diagonal blocks decreases, and vice versa, for fixed values of $p_1$, $p_2$, and $p_{12}$. In this section, we focus on how the SRD statistics evolve with varying values of intermediate $\gamma$. Specifically, we determine the range of $\gamma$ over which the SRD exhibits statistics corresponding to $m=2$ (two GOEs), and the values beyond which it transitions to $m=1$ (a single GOE).

\begin{figure*}[!ht]
    \centering
    \includegraphics[width=0.95\textwidth]{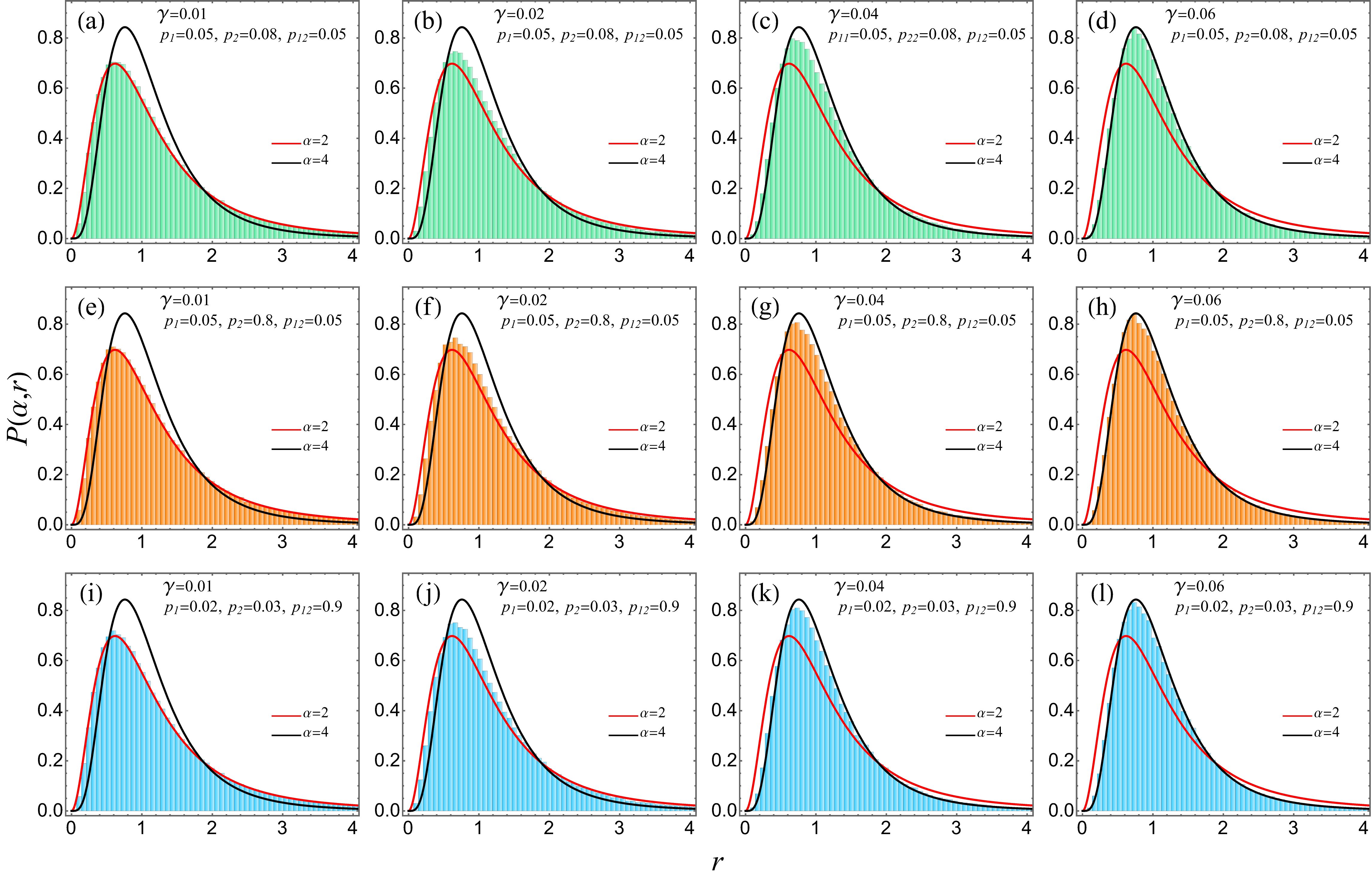}
    \caption{Second order ($k=2$) SRD plots (histogram) for bilayer network ($m=2$)  for our crossover model that is mentioned in Sec.~\ref{sublevel:level3.4}. The diagonal blocks of the adjacency matrix of model are of dimension $350 \times 350$ and $450 \times 450$, respectively. The histograms have been obtained using simulations based on an ensemble comprising 250 adjacency matrices. The crossover parameter and edge probabilities $(\gamma, p_1,p_2,p_{12})$ for the histograms obtained from numerical simulations are (a) (0.01,0.05,0.08,0.05), (b) (0.02,0.05,0.08,0.05), (c) (0.04,0.05,0.08,0.05), (d) (0.06,0.05,0.08,0.05), (e) (0.01,0.05,0.8,0.05), (f) (0.02,0.05,0.8,0.05), (g) (0.04,0.05,0.8,0.05), (h) (0.06,0.05,0.8,0.05), (i) (0.01,0.02,0.03,0.9), (j) (0.02,0.02,0.03,0.9), (k) (0.04,0.02,0.03,0.9), and (l) (0.06,0.02,0.03,0.9)}
    \label{fig:Bilayercrossover}
\end{figure*}

In Fig.~\ref{fig:Bilayercrossover}, we investigate the second-order ($k = 2$) SRD for our crossover model. The (scaled) blocks are $A_{350 \times 350}^{1}$, $A_{450 \times 450}^{2}$, and $B_{350 \times 450}^{1,2}$, and we vary the crossover parameter $\gamma$. When $\gamma = 0.01$, the system effectively exhibits a two block-diagonal structure ($m=2$), with spectral fluctuations consistent with two weakly coupled GOE ensembles, in line with Table~\ref{table:1}, as the histogram aligns closely with the $\alpha = 2$ result from RMT. As we gradually increase the value of $\gamma$, we observe intermediate statistics indicating a transition from two GOEs to a single GOE. At $\gamma=0.06$, the SRD histogram aligns closely with the $\alpha=4$ result from RMT, corresponding to a single GOE ($m=1$), as indicated by Table~\ref{table:1}. Furthermore, we find that the crossover behavior is largely insensitive to the intralayer and interlayer edge connection probabilities, provided the network dimensions are fixed. For example, as shown in Fig.~\ref{fig:Bilayercrossover}~(a), (e), and (i), the statistics remain consistent for $\gamma = 0.01$, despite different edge connection probabilities. Additionally, we examined the third-order ($k=3$) SRD for the same dimensions and observed a similar transition in SRD statistics, from two GOEs to a single GOE, as $\gamma$ is varied.

\begin{figure*}[!ht]
    \centering
    \includegraphics[width=0.95\textwidth]{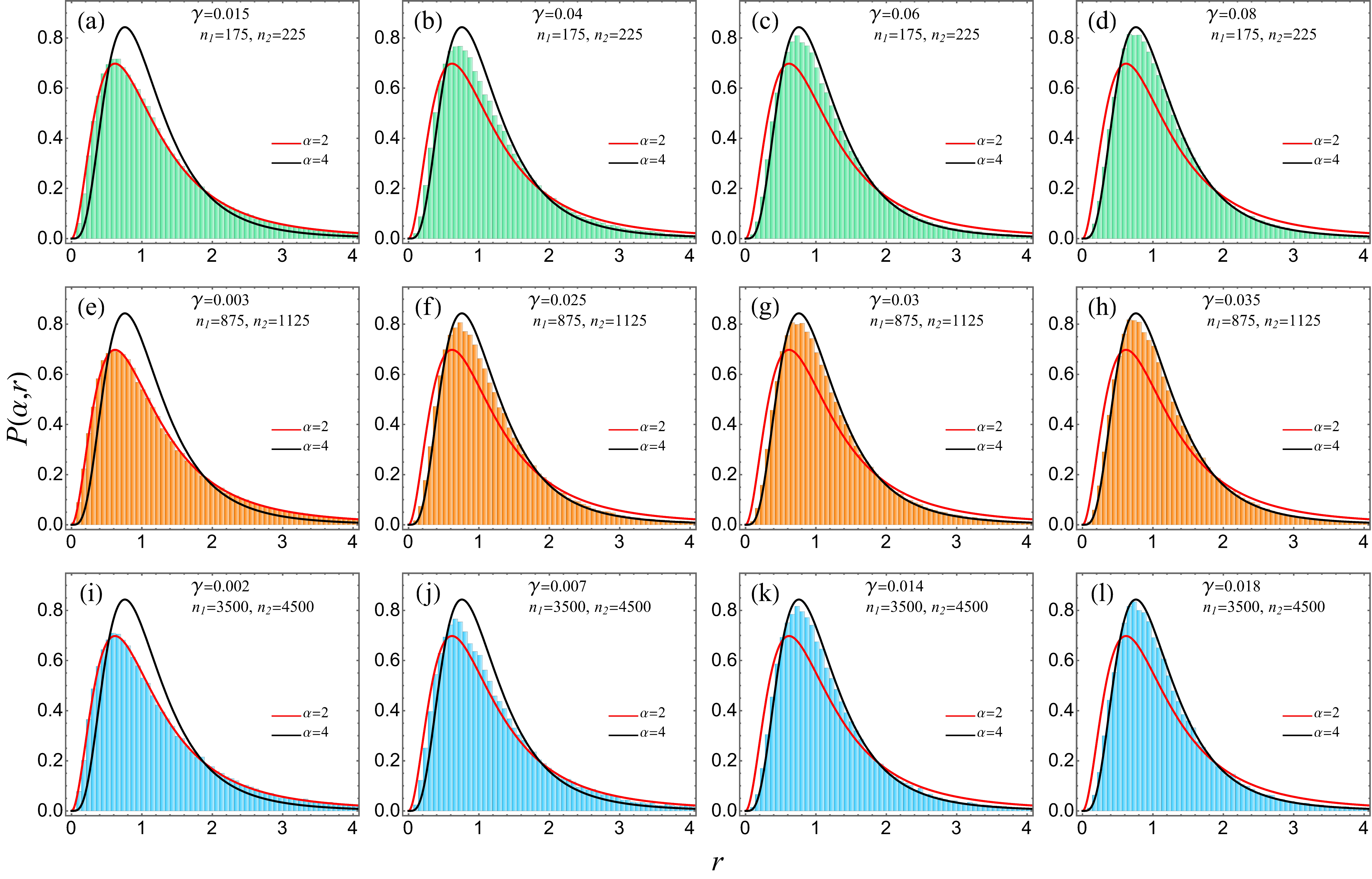}
    \caption{Second-order ($k=2$) SRD plots (histograms) for the bilayer network ($m=2$) corresponding to the crossover model introduced in Sec. ~\ref{sublevel:level3.4}. The intralayer and interlayer edge connection probabilities are fixed at $p_1 = 0.05$, $p_2 = 0.8$, and $p_{12} = 0.05$ for all Cases. The histograms are obtained from numerical simulations using ensembles of adjacency matrices. The block dimension factor $\kappa$, crossover parameter and diagonal block dimensions of the adjacency matrix, $(\kappa, \gamma, n_1, n_2)$, corresponding to the histograms are: (a)(1, 0.015, 175, 225), (b) (1, 0.04, 175, 225), (c) (1, 0.06, 175, 225),(d) (1, 0.08, 175, 225), (e) (5, 0.003, 875, 1125), (f) (5, 0.025, 875, 1125), (g) (5, 0.03, 875, 1125), (h) (5, 0.035, 875, 1125), (i) (20, 0.002, 3500, 4500), (j) (20, 0.007, 3500, 4500), (k) (20, 0.014, 3500, 4500), and (l) (20, 0.018, 3500, 4500).}
    \label{fig:alldimBilayercrossover}
\end{figure*}

We further investigated the size dependence of the crossover behavior by systematically increasing the matrix dimensions while keeping the edge connection probabilities fixed. To achieve this, we scaled the block dimensions by a factor $\kappa$, such that the block sizes transform as $n_1 \rightarrow \kappa n_1$ and $n_2 \rightarrow \kappa n_2$. This procedure allows us to increase the total system size while preserving the relative structure of the bilayer network. In our earlier analysis, we observed that for a fixed dimension, the crossover behavior remains largely insensitive to the intralayer and interlayer edge connection probabilities. Therefore, in the present analysis, we fix the probabilities at $p_1 = 0.05$, $p_2 = 0.8$, and $p_{12} = 0.05$, and vary only the system size through $\kappa$. We begin with a base system defined by $n_1 = 175$ and $n_2 = 225$ (corresponding to $\kappa = 1$). We then consider larger system sizes by increasing $\kappa$, for example $\kappa = 5$ and $\kappa = 20$, which correspond to total matrix dimensions of 2000 ($n_1 = 875$, $n_2 = 1125$) and 8000 ($n_1 = 3500$, $n_2 = 4500$), respectively. For each system size, we estimate the minimum value of the crossover parameter $\gamma$ at which the spectral statistics transition from statistics consistent with two independent GOE ensembles to those of a single GOE ensemble. We find that the required $\gamma$ decreases systematically with increasing system size. Fig.~\ref{fig:alldimBilayercrossover}(a)-(l) shows the second-order ($k=2$) SRD for different system sizes corresponding to $\kappa = 1, 5,$ and $20$, illustrating how the crossover behavior evolves with increasing system size. Specifically, the minimum $\gamma$ required is approximately $0.08$ for $\kappa = 1$, $0.035$ for $\kappa = 5$, and $0.018$ for $\kappa = 20$.

These results show that as the system dimension increases, progressively smaller interlayer coupling is sufficient to drive the spectral statistics from those of two independent GOEs to a single GOE. In particular, even for very small but finite $\gamma > 0$, the spectral fluctuations tend to follow single GOE statistics for sufficiently large systems. Alongside this, the crossover itself becomes sharper with increasing dimension: the range of $\gamma$ over which the system exhibits intermediate statistics, narrows as system size grows, so that the transition resembles an increasingly abrupt step. Taken together, these observations suggest that in the thermodynamic (infinite-dimension) limit, arbitrarily weak interlayer coupling may suffice to induce global spectral correlations, with the system transitioning almost discontinuously from an independent-layer to collective spectral behavior.

However, obtaining a universal scaling form for the crossover remains technically challenging. A natural candidate for the scaling variable might be the product $\gamma N$, where $N$ denotes the total matrix dimension. Nevertheless, the crossover threshold in $\gamma$ does not exhibit a simple scaling behavior with $N$ alone, as it is also influenced by additional structural parameters, including the ratio of block sizes $n_{1}/n_{2}$ and the intra- and interlayer connection probabilities. Consequently, defining a unique and universal scaling variable is nontrivial, and we highlight this as an important direction for future investigation.

\section{\label{sec:level4}Application of Spectral fluctuations of random multilayer network}

In this section, we explore the application of a multilayer network approach to analyze the complexity of protein structures~\cite{AJalan2014,RF2009}. When modeling proteins as networks, a node can represent an atom, residue, or structural element, while the edges (connections) between these nodes signify various types of interactions, such as covalent bonds, hydrogen bonds, or spatial proximity~\cite{AZ2009}.

In our study, we employed the distance matrix to construct an  adjacency matrix, interpreting the spatial distances between atoms or residues as a measure of connectivity or interaction strength. This approach proves valuable in structural and functional analyses of proteins as it highlights how the three-dimensional arrangement of a protein influences its interactions and also reveals the universal features captured by RMT. To construct the adjacency matrix from the distance matrix, a threshold is typically used to determine whether two node atoms (or residues) are considered adjacent. Specifically, if the distance between two atoms falls below a predefined threshold, they are regarded as adjacent, and the corresponding adjacency matrix entry $a_{xy}$ is assigned a value of 1; otherwise, they are considered not connected and $a_{xy}$ is set to 0. 

We study the spectral fluctuation statistics of interatomic distance networks in protein crystals for two different species within the framework of RMT.

(a) We consider two proteins (1EWT and 1EWK) of \textit{Rattus norvegicus}, which exhibit the global stoichiometry characteristic of a homodimer (homo 2-mer). While both are X-ray crystal structures of the metabotropic glutamate receptor subtype 1, 1EWT represents the free form (not bound to any ligand or small molecule), and 1EWK is the protein complexed with glutamate (a small molecule). Both proteins have a deposited residue count of 980. However, the number of modeled residues differs: 912 residues are modeled for 1EWT, while 897 residues are modeled for 1EWK.

(b) The 1UW6 protein from \textit{Lymnaea stagnalis} exhibits cyclic global symmetry and global stoichiometry consistent with a pentameric assembly (Homo 5-mer) and X-ray structure of acetylcholine binding receptor. It has a deposited residue count of 4,220, while the number of modeled residues is 4,134.

To analyze the fluctuation properties in protein structures, we employed the SRD alongside the cumulative SRD (CSRD). The explicit formulas for the CSRD are provided in Appendix~\ref{AppenExpli} for different $\alpha$ values, which are used in our analysis. In Sec~\ref{sec:level2}, we discussed the scaling factors for the diagonal and off-diagonal terms of the adjacency matrix of a multilayer network in terms of network dimension and edge connection probabilities. However, in real networks, the number of edges is typically known. Therefore, we can explicitly redefine the scaling factors in terms of the number of edges and the dimensions of the blocks. The scaling factor for a diagonal block is given by $1/\sqrt{\frac{8n^{j}}{n_{j}-1}\left(1-\frac{2n^{j}}{n_{j}(n_{j}-1)}\right)}$, and $1/\sqrt{\frac{4n^{jk}}{\sqrt{n_{j}n_{k}}}\left(1-\frac{n^{jk}}{n_{j}n_{k}}\right)}$ for an off-diagonal block, where $n^{j}$ denotes the total number of edges in the $j$-th diagonal block $A_{n_j\times n_j}^j$, and $n^{jk}$ denotes the total number of edges in the off-diagonal block $B_{n_j\times n_k}^{j,k}$. In this framework, the quantity $(n^j)$ depends on the threshold distance ($\text{Td}$) between the residues within a single diagonal block, representing intralayer connections. In contrast, $(n^{jk})$ is determined by the threshold distance ($\text{Td}_{\text{Inter}}$), which characterizes the interaction between residues from two different diagonal blocks, corresponding to interlayer connections.

We now examine how the threshold distance influences connectivity in protein crystal structures. Two approaches are considered for varying the threshold distance. In the first approach, the intralayer and interlayer threshold distances, denoted as $\text{Td}$ and $\text{Td}_{\text{Inter}}$, are taken to be equal and increased simultaneously, allowing us to track the evolution of the network under uniform enhancement of connectivity. In the second approach, we fix the intralayer threshold distance within each diagonal block and increase the interlayer threshold distance $\text{Td}_{\text{Inter}}$ independently, thereby selectively enhancing inter-monomer connections. These two methods provide insight into how different connectivity modes influence the structural properties and the potential transition between universal regimes in these networks.

The threshold distance is chosen such that the resulting protein contact network captures physically meaningful structural interactions. For very small threshold values, the network is highly fragmented and does not reflect global connectivity. Therefore, we start from the smallest threshold distance at which either no or only very few inter-monomer connections are present, while each individual monomer layer remains internally well connected. We then systematically increase the threshold distance to study how inter-monomer coupling influences spectral fluctuations and crossover behavior.

\begin{figure*}[!ht]
    \centering
    \includegraphics[width=0.95\textwidth]{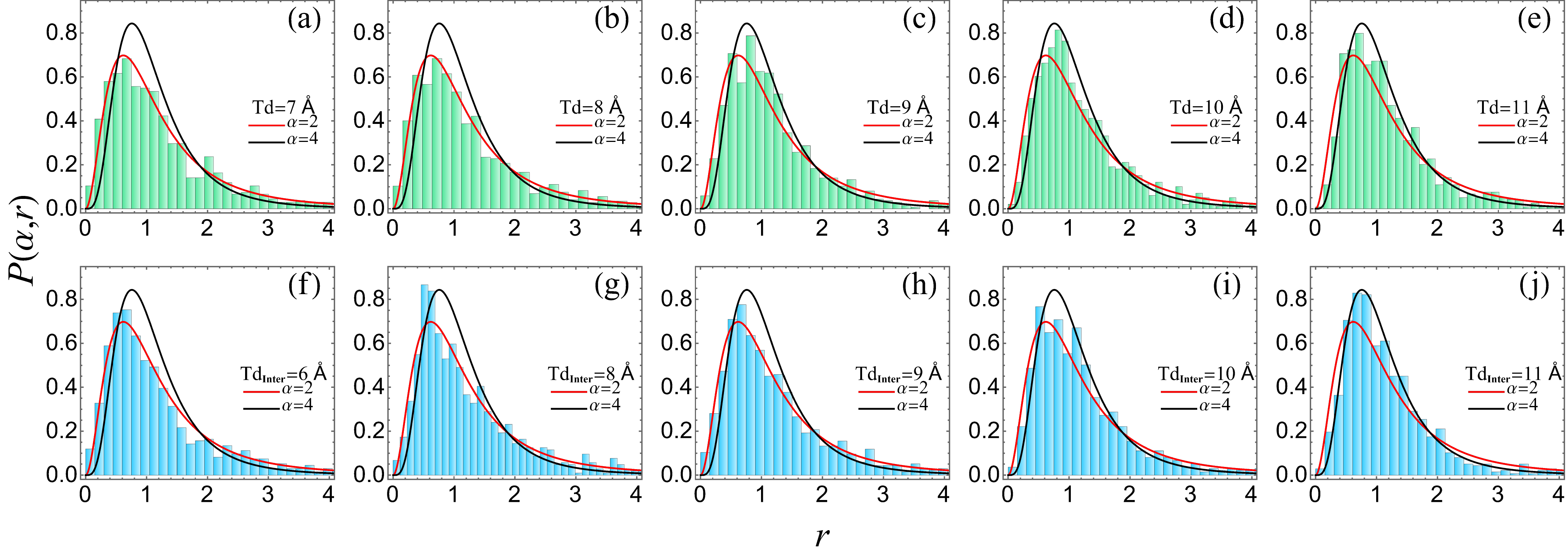}
    \caption{Second-order ($k=2$) SRD plots (histogram)  are presented for the bilayer 1EWT protein crystal structure ($m = 2$), where each diagonal block of the adjacency matrix $\mathcal{A}$ has dimensions $456 \times 456$. The histograms correspond to different combinations of intralayer and interlayer threshold distances $(\text{Td}$ and $\text{Td}_{\text{Inter}})$ in \AA , along with the number of intralayer edges in block one ($n^1$), block two ($n^2$), and the number of interlayer edges ($n^{12}$). The SRD plots are obtained for the following Cases  ($\text{Td},  \text{Td}_{\text{Inter}}, n^1, n^2, n^{12}$): (a) (7,7,1817,1826,4), (b) (8,8,2284,2274,22), (c) (9,9,3073,3052,37), (d) (10,10,4225,4227,75), (e) (11,11,5621,5597,147), (f) (7,6,1250,1248,1), (g) (7,8,1817,1826,22), (h) (7,9,1817,1826,37), (i) (7,10,1817,1826,75), and (j) (7,11,1817,1826,147).}
    \label{fig:ewt_ratio_pdf}
\end{figure*}

\begin{figure*}[!ht]
    \centering
    \includegraphics[width=0.95\textwidth]{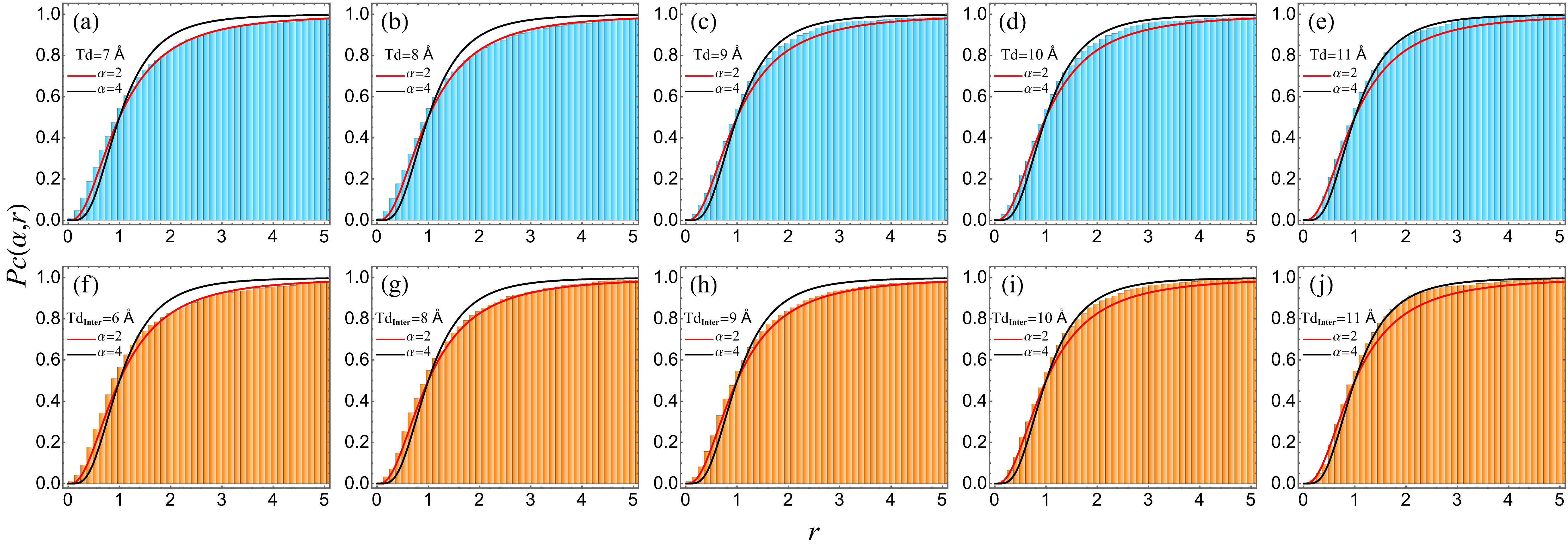}
    \caption{Second--order ($k=2$) CSRD plots (histogram), The parameter values used  as  Fig.~\ref{fig:ewt_ratio_pdf}.}
    \label{fig:ewt_ratio_cdf}
\end{figure*}

First, we analyze the protein structure 1EWT, in which each monomer has 456 residues. As a result, the corresponding adjacency matrix $\mathcal{A}$ is of dimension $912 \times 912$, with each diagonal block representing intra-monomer interactions of size $456 \times 456$, and the off-diagonal blocks corresponding to inter-monomer interactions also of size $456 \times 456$. Following the first approach, for a threshold distance of $\text{Td} = \text{Td}_{\text{Inter}} = 5.5$~\AA, block one contains 915 connections and block two contains 911 connections, with no interlayer connections. Increasing the threshold to $\text{Td}=\text{Td}_{\text{Inter}} = 6$~\AA\ introduces the first interlayer connection, while the intralayer complexity increases, yielding 1250 and 1248 connections in block one and block two, respectively. At $\text{Td}=\text{Td}_{\text{Inter}}  = 7$~\AA, four interlayer connections emerge, and the intralayer connections increase to 1817 and 1826. At this point, the second-order ($k=2$) SRD and CSRD statistics align well with $\alpha = 2$, as shown in Figs.~\ref{fig:ewt_ratio_pdf}(a) and \ref{fig:ewt_ratio_cdf}(a). Further increasing the threshold distance to $\text{Td} = \text{Td}_{\text{Inter}} = 8$~\AA\ results in 22 interlayer connections, with 2284 and 2274 intralayer connections in the two blocks. The spectral statistics remain consistent with $\alpha = 2$, as seen in Figs.~\ref{fig:ewt_ratio_pdf}(b) and \ref{fig:ewt_ratio_cdf}(b). When $\text{Td}=\text{Td}_{\text{Inter}}$ is raised to 9~\AA, 37 interlayer connections are observed, and the intralayer connections grow to 3073 and 3052 in blocks one and two, respectively. The corresponding SRD and CSRD statistics exhibit a transition toward an intermediate regime between $\alpha = 2$ and $\alpha = 4$, as shown in Figs.~\ref{fig:ewt_ratio_pdf}(c) and \ref{fig:ewt_ratio_cdf}(c). At $\text{Td}=\text{Td}_{\text{Inter}}=   10$~\AA, the number of interlayer connections increases to 75, with blocks one and two containing 4225 and 4227 intralayer connections, respectively. The statistics continue to exhibit  an intermediate regime between $\alpha = 2$ and $\alpha = 4$, as shown in Figs.~\ref{fig:ewt_ratio_pdf}(d) and \ref{fig:ewt_ratio_cdf}(d). At $\text{Td}=\text{Td}_{\text{Inter}}=   11$~\AA, the number of interlayer connections increases to 147, with block one and block two containing 5621 and 5597 intralayer connections, respectively. In this Case, the spectral statistics lie close to $\alpha = 4$, as demonstrated in Figs.~\ref{fig:ewt_ratio_pdf}(e) and \ref{fig:ewt_ratio_cdf}(e).

In the second approach, we fix the intralayer threshold distance at $\text{Td} = 7$~\AA\ and vary the interlayer threshold distance $\text{Td}_{\text{Inter}}$ from 6~\AA\ to 11~\AA. This leads to a continuous crossover in the SRD and CSRD statistics, evolving from $\alpha = 2$ to $\alpha = 4$, as shown in Figs.~\ref{fig:ewt_ratio_pdf}(f)--(j) and~\ref{fig:ewt_ratio_cdf}(f)--(j). These observations demonstrate that increasing interlayer connectivity drives the transition between universality classes.

\begin{figure*}[!ht]
    \centering
    \includegraphics[width=0.95\textwidth]{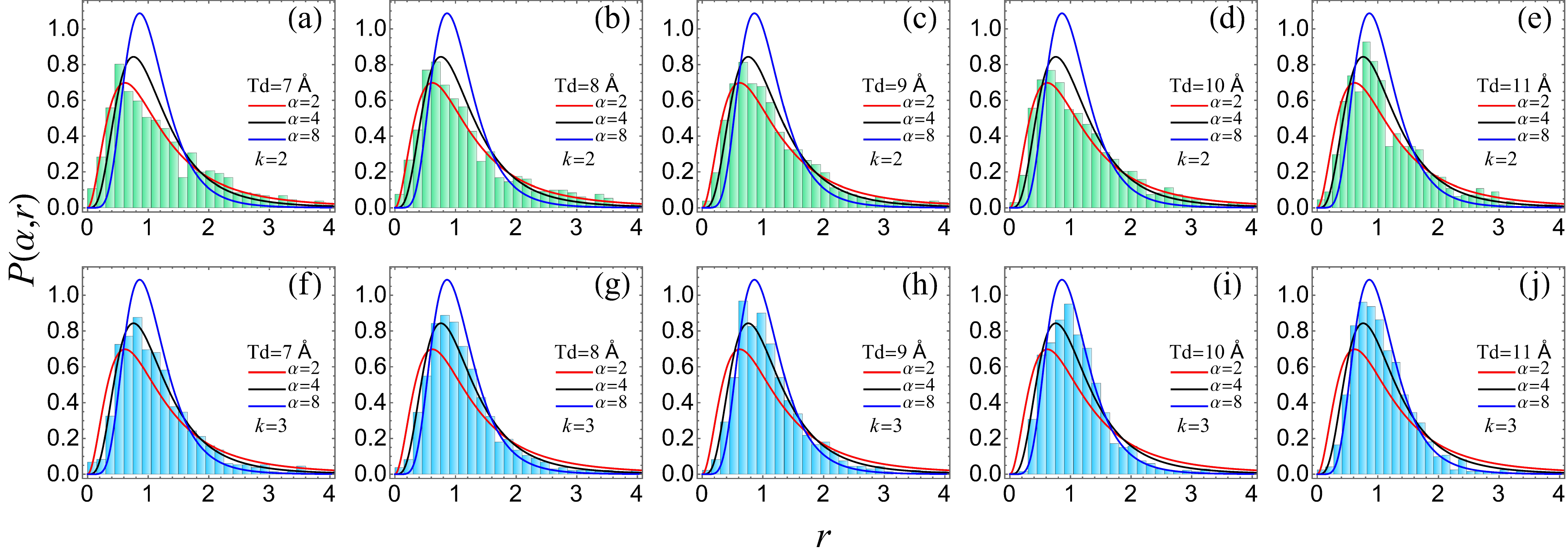}
    \caption{Higher order SRD plots (histogram) are presented for the bilayer 1EWK protein crystal structure ($m = 2$), where the diagonal blocks of the adjacency matrix $\mathcal{A}$ have dimensions $448 \times 448$ and $447 \times 447$, respectively. The histograms for second-order ($k=2$) and third-order ($k=3$) SRD correspond to different combinations of intra layer and inter layer threshold distances $(\text{Td}$ and $\text{Td}_{\text{Inter}})$ in \AA, along with the number of intra layer edges in block one ($n^1$), block two ($n^2$), and the number of interlayer edges ($n^{12}$). The SRD plots are obtained for the following Cases, denoted as $(k, \text{Td}, \text{Td}_{\text{Inter}}, n^1, n^2, n^{12})$: (a) (2,7,7,1799,1802,18), (b) (2,8,8,2267,2248,25), (c) (2,9,9,3058,3041,52), (d) (2,10,10,4122,4167,95), (e) (2,11,11,5668,5601,171), (f) (3,7,7,1799,1802,18), (g) (3,8,8,2267,2248,25), (h) (3,9,9,3058,3041,52), (i) (3,10,10,4122,4167,95), and (j) (3,11,11,5668,5601,171).}
    \label{fig:ewk_ratio_pdf}
\end{figure*}

\begin{figure*}[!ht]
    \centering
    \includegraphics[width=0.95\textwidth]{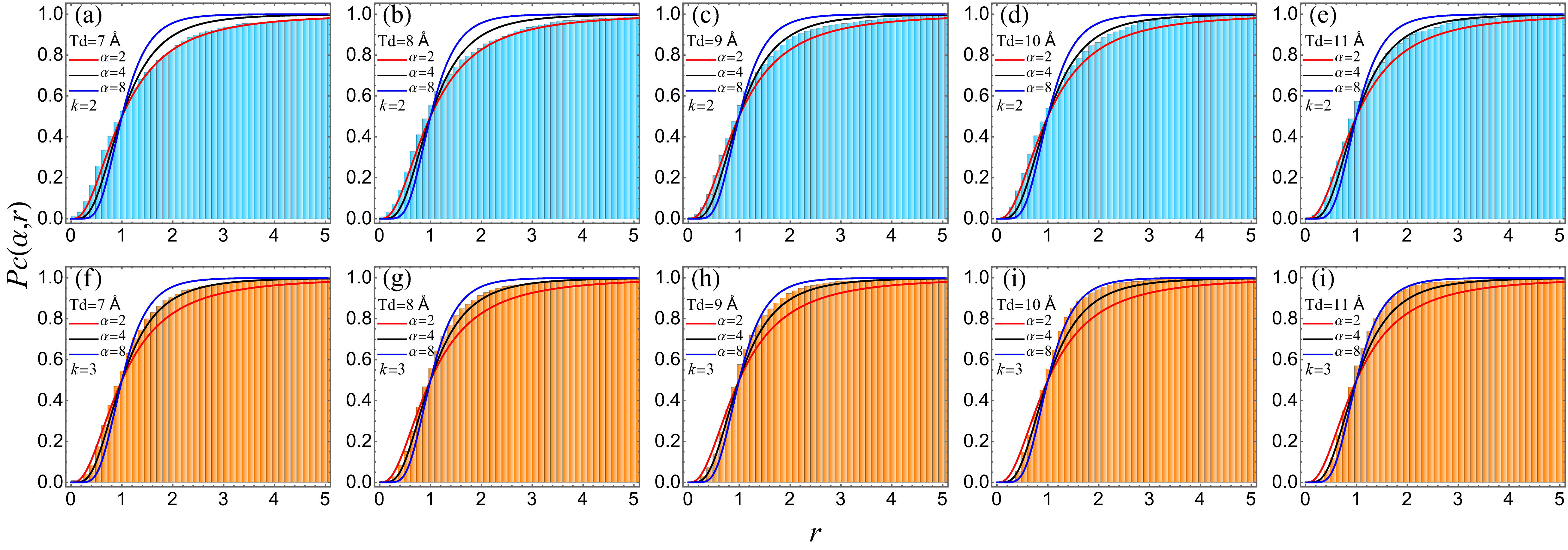}
    \caption{Highe--order of  CSRD plots (histogram), The parameter values are same as those in Fig.~\ref{fig:ewk_ratio_pdf}.}
    \label{fig:ewk_ratio_cdf}
\end{figure*}

Next, we analyze the bilayer structure of the 1EWK protein, where one monomer contains 449 residues and the other contains 450 residues. Consequently, the diagonal blocks of the adjacency matrix $\mathcal{A}$ have dimensions $449 \times 449$ and $450 \times 450$ respectively, yielding a total matrix dimension of $899 \times 899$. The upper off-diagonal block, representing interlayer connections, has dimensions $449 \times 450$. In this configuration, for the first approach, we perform the analysis by increasing both threshold parameters, $\text{Td}$ and $\text{Td}_{\text{Inter}}$, simultaneously for $k = 2$ and $k = 3$. At an initial threshold distance of $\text{Td} = \text{Td}_{\text{Inter}} = 5.5$~\AA, block one contains 931 connections, block two has 929 connections, and no interlayer connections are observed. Increasing the threshold to $\text{Td} = \text{Td}_{\text{Inter}} = 6$~\AA, five interlayer connections emerge, while intralayer complexity increases, with block one and block two containing 1246 and 1244 connections, respectively. At $\text{Td} = \text{Td}_{\text{Inter}} = 7$~\AA, 18 interlayer connections are observed, with 1799 and 1802 intralayer connections in blocks one and two, respectively. At this threshold, the statistics of the second-order ($k=2$) SRD and CSRD align well with $\alpha = 2$, as shown in Figs.~\ref{fig:ewk_ratio_pdf}(a) and \ref{fig:ewk_ratio_cdf}(a). Further increasing the threshold to $\text{Td} = \text{Td}_{\text{Inter}} = 8$~\AA leads to a rise in interlayer connections, with block one containing 2267 and block two containing 2248 intralayer connections. The SRD and CSRD statistics remain close to the $\alpha = 2$ result, as shown in Figs.~\ref{fig:ewk_ratio_pdf}(b) and \ref{fig:ewk_ratio_cdf}(b). At $\text{Td} = \text{Td}_{\text{Inter}} = 9$~\AA, 52 interlayer connections emerge, and intralayer connections increase to 3058 and 3041 in the respective blocks. For $k=2$, SRD and CSRD statistics shift to an intermediate regime between $\alpha = 2$ and $\alpha = 4$, as illustrated in Figs.~\ref{fig:ewk_ratio_pdf}(c) and \ref{fig:ewk_ratio_cdf}(c). At $\text{Td} = \text{Td}_{\text{Inter}} = 10$~\AA, the number of interlayer edges rises to 95, with intralayer edges totaling 4122 and 4167, and the statistics remain in the intermediate regime between $\alpha = 2$ and $\alpha = 4$, as seen in Figs.~\ref{fig:ewk_ratio_pdf}(d) and \ref{fig:ewk_ratio_cdf}(d). Finally, at $\text{Td} = \text{Td}_{\text{Inter}} = 11$~\AA, 171 interlayer connections are observed, and the intralayer edges increase to 5668 and 5601. In this Case, the SRD and CSRD statistics for $k=2$ closely follow the distribution associated with $\alpha = 4$, as shown in Figs.~\ref{fig:ewk_ratio_pdf}(e) and \ref{fig:ewk_ratio_cdf}(e). Additionally, for the third-order ($k=3$) Case, the SRD and CSRD statistics display a continuous crossover from $\alpha = 4$ to $\alpha = 8$, as illustrated in Figs.~\ref{fig:ewk_ratio_pdf}(f)--(j) and \ref{fig:ewk_ratio_cdf}(f)--(j).

In the second approach, we again fix the intralayer threshold distance while varying the interlayer threshold distance independently. This yields similar statistical behavior, with a continuous crossover in the SRD and CSRD statistics. For $k=2$, the statistics evolve from $\alpha = 2$ to $\alpha = 4$, while for $k=3$, they evolve from $\alpha = 4$ to $\alpha = 8$, indicating a gradual transition in the underlying structural correlations.

\begin{figure*}[!ht]
    \centering
    \includegraphics[width=0.95\textwidth]{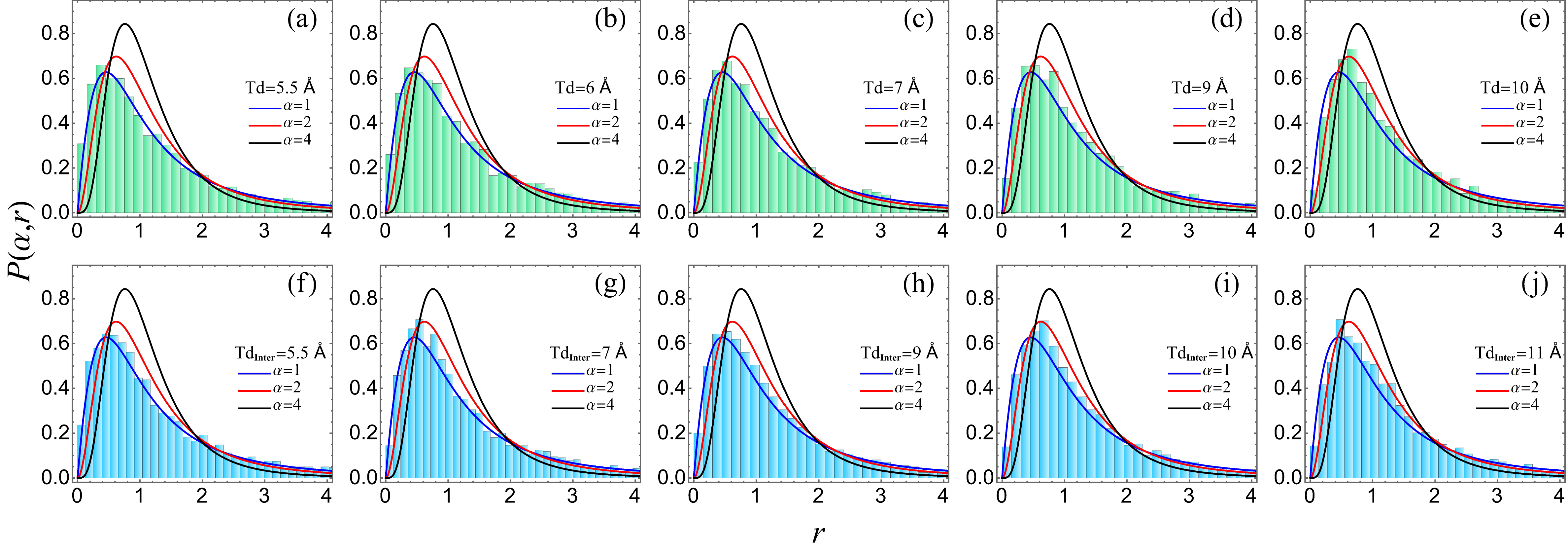}
    \caption{Second--order ($k=2$) SRD plots (histogram) are presented for the tetralayer 1UW6 protein crystal structure ($m=4$). The diagonal blocks of the adjacency matrix $\mathcal{A}$ correspond to dimensions $1034\times 1034$, $1034\times 1034$, $1033\times 1033$, and $1033\times 1033$ for blocks one through four, respectively. The histograms represent various combinations of intralayer and interlayer threshold distances $(\text{Td}, \text{Td}_{\text{Inter}})$ in \AA, along with the number of intralayer edges in block one ($n^1$), block two ($n^2$), block three ($n^3$), block four ($n^4$), and the number of interlayer edges between block pairs: one and two ($n^{12}$), one and three ($n^{13}$), one and four ($n^{14}$), two and three ($n^{23}$), two and four ($n^{24}$), and three and four ($n^{34}$). The SRD plots are obtained for the following parameter sets: (a) (5.5,5.5,1852,1862,1873,1888,1,0,0,0,1,0), (b) (6,6,2651,2645,2656,2663,1,0,0,0,3,0), (c) (7,7,4101,4111,4094,4086,5,0,0,0,4,0), (d) (9,9,6932,6931,6933,6950,25,0,0,0,39,0), (e) (10,10,9632,9602,9630,9618,58,0,0,0,70,0), (f) (6,6,2651,2645,2656,2663,1,0, 0,0,1,0), (g) (6,7,2651,2645,2656,2663,5,0,0,0,4,0), (h) (6,9,2651,2645,2656,2663,25,0,0,0,39,0), (i) (6,10,2651,2645,2656,2663,58,0,0,70), and (j) (6,11,2651,2645,2656,2663,95,0,0,0,113).}
    \label{fig:1uw6_ratio_pdf}
\end{figure*}

\begin{figure*}[!ht]
    \centering
    \includegraphics[width=0.95\textwidth]{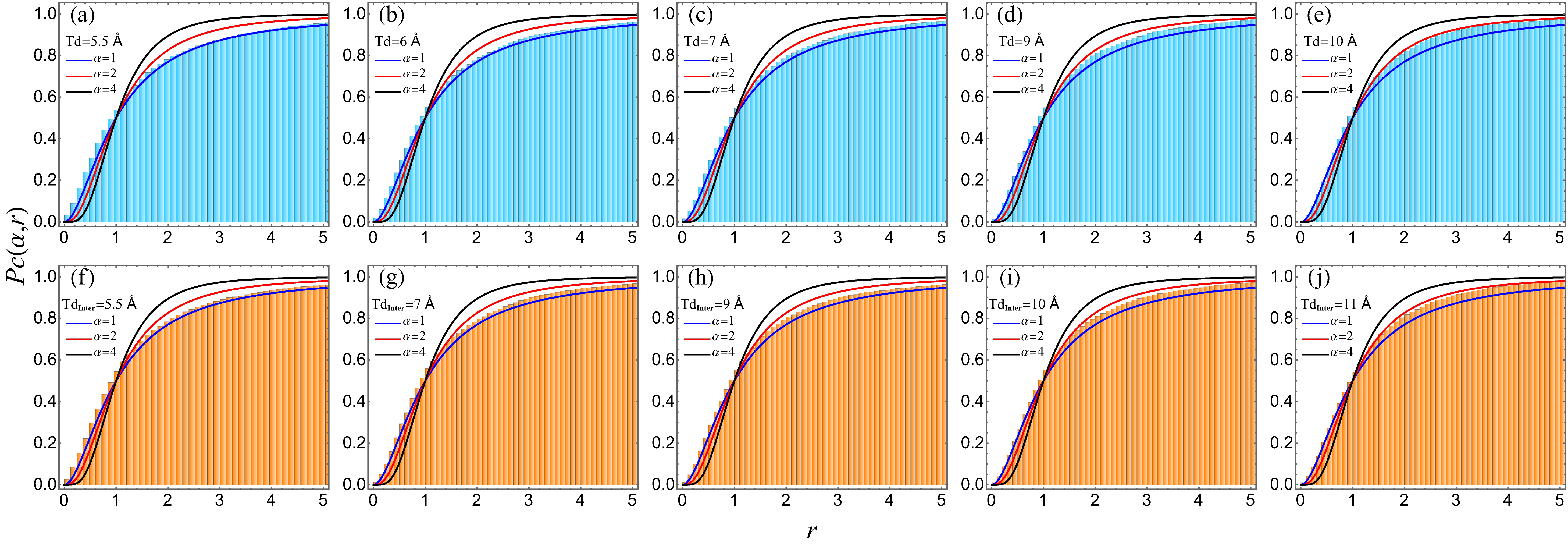}
    \caption{Second--order ($k=2$) CSRD plots (histogram), The parameter values used  as Fig.~\ref{fig:1uw6_ratio_pdf}}
    \label{fig:1uw6_ratio_cdf}
\end{figure*}

Finally, in the structure 1UW6, five monomers assemble to form a pentameric unit. Subsequently, four such pentameric units are further organized into a Higher--order quaternary structure, resulting in a 20-subunit capsid-like assembly. In this protein crystal structure, we consider 4134 residues involved in interactions, resulting in an adjacency matrix $\mathcal{A}$ of dimensions $4134 \times 4134$. The residues are divided into four block structures, with blocks one and two containing 1033 residues each, and blocks three and four containing 1034 residues each. Initially, in the first approach, at a threshold distance $\text{Td} = \text{Td}_{\text{Inter}} = 5$~\AA, the number of edges in blocks one, two, three, and four are 1349, 1350, 1352, and 1348, respectively, with a single interlayer connection between blocks one and two. At $\text{Td} = \text{Td}_{\text{Inter}} = 5.5$~\AA, the number of edges increases to 1852, 1862, 1873, and 1888 in blocks one through four, respectively, with two interlayer connections: one between blocks one and two, and another between blocks two and four. The second-order ($k=2$) SRD and CSRD statistics align with $\alpha = 1$, as shown in Figs.~\ref{fig:1uw6_ratio_pdf}(a) and \ref{fig:1uw6_ratio_cdf}(a), indicating the presence of a well-defined four-block network structure, in agreement with Table~\ref{table:1}. As the threshold distance increases to $\text{Td} = \text{Td}_{\text{Inter}} = 6$~\AA, the edge count grows to 2651, 2645, 2656, and 2663 in the respective blocks, with four interlayer connections: one between blocks one and two, and three between blocks two and four. The SRD and CSRD statistics continue to exhibit agreement with $\alpha=1$ [Figs.~\ref{fig:1uw6_ratio_pdf}(b)] and \ref{fig:1uw6_ratio_cdf}(b)). At $\text{Td} = \text{Td}_{\text{Inter}} = 7$~\AA, the number of edges rises to 4101, 4111, 4014, and 4086, respectively. Nine interlayer connections are observed—five between blocks one and two, and four between blocks two and four resulting in SRD and CSRD statistics that show intermediate behavior between $\alpha = 1$ and $\alpha = 2$ [Figs.~\ref{fig:1uw6_ratio_pdf}(c)] and \ref{fig:1uw6_ratio_cdf}(c)). When the threshold is increased to $\text{Td} = \text{Td}_{\text{Inter}} = 8$~\AA, there are 5187, 5174, 5370, and 5190 intralayer connections respectively with 29 interlayer connections (13 between blocks one and two, and 16 between blocks two and four). At $\text{Td} = \text{Td}_{\text{Inter}} = 9$~\AA, the intralayer edge counts increase to 6932, 6931, 6933, and 6950 respectively, with 64 interlayer connections (25 between blocks one and two, and 39 between blocks two and four), the SRD and CSRD statistics remains intermediate between $\alpha = 1$ and $\alpha = 2$ [Figs.~\ref{fig:1uw6_ratio_pdf}(d) and \ref{fig:1uw6_ratio_cdf}(d)]. Finally, at $\text{Td} = \text{Td}_{\text{Inter}} = 10$~\AA, the number of intralayer connections reach 9632, 9602, 9630, and 9618 respectively, with 138 interlayer connections (58 between blocks one and two, and 70 between blocks two and four), while block three remains isolated. At this stage, the second-order ($k=2$) SRD and CSRD statistics transition fully to $\alpha = 2$, indicating that the original four-block structure has effectively merged into a two-block system, where blocks one, two, and four form a single block through interlayer interactions, while block three remains uncoupled [Figs.~\ref{fig:1uw6_ratio_pdf}(e) and \ref{fig:1uw6_ratio_cdf}(e)]. Implementing the second approach to analyze crossover behavior, the intralayer threshold distance is fixed at $\text{Td} = 6$~\AA, while the interlayer threshold distance $\text{Td}_{\text{Inter}}$ is varied independently from 5.5~\AA\ to 11~\AA. This analysis reveals a continuous crossover in the SRD and CSRD statistics, evolving from $\alpha = 4$ to $\alpha = 2$, as shown in Figs.~\ref{fig:1uw6_ratio_pdf}(f)--(j) and \ref{fig:1uw6_ratio_cdf}(f)--(j).

\section{\label{sec:level5}Summary and Conclusion}

In this work, we show that variance mismatch across blocks is the central obstacle to observing RMT universality in multilayer networks and introduce a general block--wise normalization scheme that resolves this issue across arbitrary architectures. Once properly normalized, Higher--order spectral fluctuations in multilayer networks exhibit universal behavior consistent with RMT across a wide range of configurations, including purely intralayer, interlayer, and multiplex structures, establishing universality as a robust feature of multilayer network spectra.

We introduced a crossover model for bilayer networks parametrized by the relative interlayer to intralayer coupling strength $\gamma$, which captures the continuous transition in spectral statistics from two independent GOEs to a single GOE. The minimum $\gamma$ required for this transition decreases systematically with system size, and the crossover sharpens with increasing dimension, with the transition range narrowing as $N$ grows. This suggests that in the thermodynamic limit, arbitrarily weak interlayer coupling may suffice to induce global spectral correlations, with the system transitioning from independent-layer to collectively coupled spectral behavior.

This spectral transition has direct physical consequences for dynamical processes on multilayer networks. The crossover threshold $\gamma$ marks the point at which layers cease to behave as independent dynamical subsystems, and its size dependence implies that large multilayer systems are particularly susceptible to interlayer-driven collective behavior. Once coupling exceeds this threshold, small perturbations can propagate across layers and trigger cascading failures~\cite{Baxter2012}, while in percolation-type processes, diffusion and transport become synchronized across the full system~\cite{Gomez2013}, suggesting that the spectral crossover can serve as a quantitative signature of the onset of system-wide dynamical coherence.

Applying this framework to protein crystal structures (1EWT, 1EWK, and 1UW6), modeled as multilayer interatomic distance networks, we observed analogous crossover behavior driven by threshold-dependent connectivity. As the distance threshold increases, the spectral statistics transition from those of independent monomers to a collectively coupled system, indicating that the same RMT framework can capture the emergence of structural integration in biological assemblies. This also suggests that spectral fluctuation analysis can serve as a tool for characterizing modular organization in complex biomolecular systems, with potential applications in protein classification and drug design.

An important direction for future work is the extension to correlated network architectures, including small-world and scale-free multilayer networks \cite{Farkas2001}. Unlike Erd\H{o}s–R\'{e}nyi networks, which are characterized by independent edge formation and exhibit weak structural correlations, these systems display strong topological correlations arising from clustering, degree heterogeneity, and hub-dominated connectivity. Understanding how such correlations modify universality and crossover behavior in multilayer systems remains an open and physically relevant problem.

\section*{Acknowledgments}
H.S. and A.D. acknowledge financial support from Shiv Nadar Institution of Eminence through institutional fellowships and computational facilities.  


\section*{DATA AVAILABILITY}
DATA AVAILABILITY
The data that support the findings of this article are openly available \cite{data1, data2, data3}.

\begin{widetext}
\appendix
\section{Closed-form expressions for cumulative spacing ratio distribution}
\label{AppenExpli}

The cumulative spacing ratio distribution (CSRD) is defined as
\begin{align}\label{CDF_WDsurmise}
P_{c}(\alpha,s)&=\int_{0}^{s}{P(\alpha,r)}~dr \nonumber\\
&=\int_{0}^{s}{C_{\alpha} \frac{(r+r^2)^\alpha}{(1+r+r^2)^{1+3 \alpha/2}}}~dr.
\end{align}
Here we provide explicit closed-form expressions for CSRD for different $\alpha$ values used in this work. 

For $\alpha=1$:
\begin{align}\label{CDF_1}
P_{c}(\alpha=1,s)&=\frac{1}{4} \left( 2 + \frac{(1 + 2s)(-2 + s + s^2)}{(1 + s + s^2)^{3/2}} \right),
\end{align}

$\alpha=2$:
\begin{align}\label{CDF_2}
P_{c}(\alpha=2,s)&=-\frac{1}{2} 
+ \frac{3 \sqrt{3} (-1 + s) s (1 + s) (2 + s) (1 + 2s)}{4 \pi (1 + s + s^2)^3} 
+ \frac{3 \arctan\left(\frac{1 + 2s}{\sqrt{3}}\right)}{\pi}
\end{align}

$\alpha=4$:
\begin{align}\label{CDF_3}
P_{c}(\alpha=4,s)&=-\frac{1}{2} 
+ \frac{3 \sqrt{3} (-1 + s) s (1 + s) (2 + s) (1 + 2s) 
\left(2 + s (1 + s) \big(6 + s (1 + s) \big(15 + 2s (1 + s)\big)\big)\right)}{8 \pi (1 + s + s^2)^6} \nonumber\\
&+ \frac{3 \arctan\left(\frac{1 + 2s}{\sqrt{3}}\right)}{\pi}
\end{align}

$\alpha=8$:
\begin{align}\label{CDF_4}
P_{c}(\alpha=8,s) &= -\frac{1}{2} 
+ \frac{3 \sqrt{3} \, (-1 + s) \, s \, (1 + s) \, (2 + s) \, (1 + 2s)}{560 \pi (1 + s + s^2)^{12}} \nonumber \\
&\quad \times \left(140 + s(1 + s) \left(1260 + s(1 + s) \left(5670 + s(1 + s) \left(15540 + s(1 + s) \right.\right.\right.\right. \nonumber \\
&\qquad \left.\left.\left.\left. \left(30492 + s(1 + s) \left(40446 + s(1 + s) \left(51099 + 14s(1 + s) \left(873 + 5s(1 + s) \left(27 + 2s(1 + s)\right) \right) \right) \right) \right) \right) \right) \right) \right)\nonumber \\
&\quad + \frac{3 \arctan\left(\frac{1 + 2s}{\sqrt{3}}\right)}{\pi}
\end{align}

\end{widetext}

\end{document}